\documentclass[%
 aip,
 amsmath,amssymb,
reprint, onecolumn 
]{revtex4-2}

\usepackage{graphicx}
\usepackage{dcolumn}
\usepackage{bm}

\usepackage[utf8]{inputenc}
\usepackage[T1]{fontenc}
\usepackage{mathptmx}
\usepackage{etoolbox}
\usepackage{color}

\usepackage[caption=false]{subfig}
\usepackage[colorlinks=true,linkcolor=blue,citecolor=red]{hyperref}

\usepackage{comment}
\usepackage{multirow}

\makeatletter
\def\@email#1#2{%
 \endgroup
 \patchcmd{\titleblock@produce}
  {\frontmatter@RRAPformat}
  {\frontmatter@RRAPformat{\produce@RRAP{*#1\href{mailto:#2}{#2}}}\frontmatter@RRAPformat}
  {}{}
}%
\makeatother

\def\pa{\partial\Omega}
\def\T{{\mathcal T}}

\def\R{{\mathbb R}}

\def\M{{\mathcal M}}

\def\P{{\mathbb P}}

\def\x{\bm{x}}
\def\y{\bm{y}}

\def\pa{{\partial \Omega}}

\def\M{\mathcal{M}}
\def\P{\mathbb{P}}

\def\R{\mathbb{R}}

\def\T{\mathcal{T}}

\def\II{\mathcal{I}}

\def\erfc{\mathrm{erfc}}
\def\erfcx{\mathrm{erfcx}}

\def\hmax{{h_{\rm max}}}
\def\nmax{{n_{\rm max}}}

\begin{document}

\preprint{AIP/123-QED}

\title[The Steklov problem for exterior domains]{The Steklov problem for exterior domains: asymptotic behavior and applications}
\author{D. S. Grebenkov}
 \email{denis.grebenkov@polytechnique.edu}
\affiliation{ 
Laboratoire de Physique de la Mati\`ere Condens\'ee, CNRS, Ecole Polytechnique, Institut Polytechnique de Paris, 91120 Palaiseau, France }

\affiliation{CNRS -- Universit\'e de Montr\'eal CRM -- CNRS,
6128 succ Centre-Ville, Montr\'eal QC H3C 3J7, Canada}

\author{A. Chaigneau}%
\affiliation{ 
Laboratoire de Physique de la Mati\`ere Condens\'ee, CNRS, Ecole Polytechnique, Institut Polytechnique de Paris, 91120 Palaiseau, France
}%

\date{\today}

\begin{abstract}
We investigate the spectral properties of the Steklov problem for the
modified Helmholtz equation $(p-\Delta) u = 0$ in the exterior of a
compact set, for which the positive parameter $p$ ensures exponential
decay of the Steklov eigenfunctions at infinity.  We obtain the
small-$p$ asymptotic behavior of the eigenvalues and eigenfunctions
and discuss their features for different space dimensions.  These
results find immediate applications to the theory of stochastic
processes and unveil the long-time asymptotic behavior of probability
densities of various first-passage times in exterior domains.
Theoretical results are validated by solving the exterior Steklov
problem by a finite-element method with a transparent boundary
condition.
\end{abstract}

\maketitle

\section{Introduction}

The Laplace operator plays the central role in mathematics, physics,
and many other disciplines
\cite{Rayleigh,Jackson,Davies,Bass,Gardiner,Sneddon,Schuss,Redner,Metzler,Lindenberg,Grebenkov,Grebenkov13}.
So, its eigenmodes, satisfying $-\Delta u = \lambda u$ in a considered
bounded domain $\Omega \subset \R^d$ with an appropriate boundary
condition, can represent vibration modes of a drum, quantum states of
a free particle in a confinement (e.g., a waveguide), or proper basis
functions for diffusive processes, whereas the Laplacian eigenvalues
are respectively related to frequencies, energies, and decay times of
these eigenmodes.  In a similar spectral problem formulated by Steklov
\cite{Steklov1902,Kuznetsov14}, the spectral parameter stands in the
Robin-type boundary condition, i.e., one searches for eigenpairs
$\{\mu, u\}$ satisfying
\begin{equation}  \label{eq:Steklov_def0}
\Delta u = 0  \quad \textrm{in}~\Omega,  \qquad  \partial_n u = \mu u  \quad \textrm{on}~\pa ,
\end{equation}
where $\partial_n$ is the normal derivative on the boundary $\pa$
oriented outwards the domain $\Omega$.  Despite intensive mathematical
studies over the past century (see the recent book \cite{Levitin} and
reviews \cite{Girouard17,Colbois24}), the Steklov problem is much less
known in physics, while its physical applications, such as electrical
impedance tomography
\cite{Cheney99,Calderon80,Borcea02,Sylvester87,Curtis91} or sloshing
problem \cite{Fox83,Kozlov04,Levitin22}, are less numerous.  Recently,
the eigenmodes of the Steklov problem have found a new application
within the encounter-based approach to diffusion-controlled reactions
\cite{Grebenkov19,Grebenkov19c,Grebenkov20,Grebenkov23c}.
In particular, the stochastic formulation of such diffusion-mediated
phenomena has naturally led to spectral expansions related to the
Steklov problem.

One of the crucial advantages of the Steklov problem manifests for
unbounded Euclidean domains with a bounded boundary.  While former
spectral expansions based on Laplacian eigenfunctions become
unavailable due to the continuum spectrum of the Laplace operator, the
spectrum of the Steklov problem is still discrete, allowing one to
employ its spectral expansions.  For instance, Auchmuty suggested
using the Steklov eigenfunctions for representing harmonic functions
in exterior domains and for solving associated boundary value problems
\cite{Auchmuty04,Auchmuty13,Auchmuty14,Auchmuty15,Auchmuty18}.  In
diffusion applications, one often deals with particles diffusing in
the unbounded space towards bounded targets
\cite{Samson77,Torquato86,Lee89,Traytak95,Galanti16,Grebenkov20a} so
that the Steklov eigenmodes can be particularly valuable (see, e.g.,
the statistics of encounters between two diffusing disks
\cite{Grebenkov21a}).

Despite this growing interest, most former works on the Steklov
problem were restricted to bounded domains.  From the mathematical
point of view, the analysis of the Steklov problem is more difficult
for exterior problems.  For instance, a constant function, which is
the principal Steklov eigenmode for any bounded domain, is not square
integrable in the case of unbounded domains.  More generally, as
formal solutions of the exterior Steklov problem may not be square
integrable, one needs either to employ larger functional spaces for a
proper formulation of the Steklov problem
\cite{Auchmuty14,Auchmuty18,Xiong23}, or to consider the interior
Steklov problem in a sequence of bounded domains approaching the
unbounded one \cite{Arendt15}.  From the numerical point of view, one
cannot directly apply common tools such as a finite-element method
(FEM) to an unbounded domain that requires a truncation of the
original domain and dealing with the resulting artificial boundary
(see below).  From the applicative point of view, the limited
knowledge on the exterior Steklov problem delays its versatile
applications in physics and other disciplines.

In this paper, we follow another approach to formulate and investigate
the exterior Steklov problem.  We start from a generalized Steklov
problem, in which the Laplace equation is replaced by the modified
Helmholtz equation, $(p - \Delta) u = 0$, with a fixed parameter $p >
0$.  An exponential decay of solutions of this equation at infinity
allows one to handle the generalized Steklov problem in exterior
domains in the same conventional way, as for bounded domains
\cite{Auchmuty13} (see also Appendix \ref{sec:upper} for
exponentially decaying upper bounds).  This spectral problem presents
its own mathematical interest and has direct applications to
stochastic processes, first-passage phenomena, and
diffusion-controlled reactions, as illustrated in
Sec. \ref{sec:application}.  We focus on the asymptotic behavior of
the eigenvalues and eigenfunctions of the generalized exterior Steklov
problem in the limit $p\to 0$.  In particular, we show how this
behavior strongly depends on the dimension $d$ of the embedding
Euclidean space $\R^d$.  The limits of the eigenvalues and
eigenfunctions at $p = 0$ are then {\it postulated} to be the
eigenmodes of the exterior Steklov problem (\ref{eq:Steklov_def0}).
The existence of these limits and the equivalence of this approach
with other formulations \cite{Auchmuty14,Arendt15} of the exterior
Steklov problem are proved in a separate work \cite{Bundrock25}.

The paper is organized as follows.  In Sec. \ref{sec:summary}, we
formulate the generalized exterior Steklov problem and recall its
basic mathematical properties.  In Sec. \ref{sec:ball}, we inspect the
exterior of a ball as a guiding example, for which this spectral
problem admits an explicit solution.  Its spectral properties bring
some intuition and practical tools for studying the general case in
Sec. \ref{sec:asympt}.  Section \ref{sec:2D} presents some numerical
illustrations for two-dimensional and three-dimensional domains.  In
Sec. \ref{sec:application}, we discuss an application of the derived
asymptotic results to diffusion-controlled reactions and the related
first-passage time statistics.  In particular, we derive the long-time
behavior of the survival probability for exterior domains in three
dimensions.  Section \ref{sec:conclusion} concludes the paper.

\section{Generalized exterior Steklov problem}
\label{sec:summary}

Let $\Omega \in \R^d$ ($d \geq 2$) be the exterior of a compact set,
and $\pa$ denote its smooth connected boundary (while the interior
Steklov problem is often formulated for domains with Lipschitz
boundary \cite{Levitin}, we deliberately avoid the related subtle
points by focusing on smooth boundaries; at the same time, numerical
examples in Sec. \ref{sec:2D} include polygonal domains; see also more
general formulations in \cite{Bundrock25}).
We search for the eigenvalues $\mu_k^{(p)}$ and eigenfunctions
$V_k^{(p)}$ of the generalized Steklov eigenvalue problem
\cite{Auchmuty13}
\begin{equation} \label{eq:Steklov}
(p - \Delta) V_k^{(p)} = 0 \quad \textrm{in}~ \Omega, \qquad \partial_n V_k^{(p)} = \mu_k^{(p)} V_k^{(p)}  \quad \textrm{on}~ \pa ,
\qquad \lim\limits_{|\x|\to \infty} V_k^{(p)} = 0 ,
\end{equation}
where $p > 0$ is a fixed parameter, $\Delta$ is the Laplace operator,
and $\partial_n$ is the normal derivative oriented outwards the domain
$\Omega$.  The positive eigenvalues $\mu_k^{(p)}$ are ordered to form
an increasing sequence that accumulates to infinity:
\begin{equation}  \label{eq:muk_ordering}
0 \leq \mu_0^{(p)} \leq \mu_1^{(p)} \leq \ldots \nearrow +\infty .
\end{equation}
We note that the restriction of $V_k^{(p)}$ onto the boundary $\pa$,
denoted as $v_k^{(p)} = V_k^{(p)}|_{\pa}$, is an eigenfunction of the
associated Dirichlet-to-Neumann operator $\M_p$ that maps a given
function $f$ on $\pa$ onto another function $g = (\partial_n
u)|_{\pa}$, where $u$ is the solution of the boundary value problem
\begin{equation}
(p-\Delta) u = 0 \quad \textrm{in}~ \Omega, \qquad  u|_{\pa} = f, \qquad \lim\limits_{|\x|\to\infty} u = 0.  
\end{equation}
As for bounded domains, the eigenfunctions $v_k^{(p)}$ form a complete
orthonormal basis of the space $L^2(\pa)$ of square-integrable
functions on the boundary $\pa$ \cite{Auchmuty13}.  In particular,
their normalization also fixes the normalization of the Steklov
eigenfunctions:
\begin{equation} \label{eq:normalization}
\int\nolimits_{\pa} |V_k^{(p)}|^2 = 1.
\end{equation}

In the conventional setting of bounded domains, solutions of
Eq. (\ref{eq:Steklov}) are searched in the space $H^1(\Omega)$ of
square-integrable functions with square-integrable derivatives.  When
$p > 0$, solutions $V_k^{(p)}$ of Eq. (\ref{eq:Steklov}) exhibit an
exponential decay as $|\x|\to \infty$ (see, e.g., upper bounds in
Appendix \ref{sec:upper}) and thus belong to the space $H^1(\Omega)$
(see \cite{Auchmuty13} for further mathematical details).  In turn,
the analysis is more subtle for $p = 0$.  A rigorous formulation of
the exterior Steklov problem in two dimensions was provided in
\cite{Christiansen23}.  In higher dimensions, one can either employ
functional spaces larger than $H^1(\Omega)$
\cite{Auchmuty14,Auchmuty18,Xiong23}, or consider the interior Steklov
problem in a sequence of bounded domains approaching $\Omega$
\cite{Arendt15}.  In the following, we assume that $\mu_k^{(p)}$ and
$V_k^{(p)}$ converge to their limits $\mu_k^{(0)}$ and $V_k^{(0)}$
that are {\it declared} to be eigenvalues and eigenfunctions of the
exterior Steklov problem at $p = 0$.  Mathematical proofs of this
convergence and the equivalence with two other formulations of the
exterior Steklov problem at $p = 0$ are beyond the scope of this paper
and will be presented in a separate work \cite{Bundrock25}.

In this paper, we aim at studying the limiting behavior of the Steklov
eigenvalues and eigenfunctions as $p\to 0$ to highlight differences
and similarities between interior and exterior Steklov problems.  In
particular, we address two questions:

(i) For a bounded domain, the principal eigenfunction $v_0^{(0)}$
corresponding to the smallest eigenvalue $\mu_0^{(0)} = 0$ is a
constant.  Does it hold for exterior domains?

(ii) For a bounded domain, the eigenvalues $\mu_k^{(p)}$ exhibit a
linear asymptotic behavior as $p\to 0$,
\begin{equation}   \label{eq:muk_small_p_bounded}
\mu_k^{(p)} = \mu_k^{(0)} + b_k\, p + o(p) ,
\end{equation}
where 
\begin{equation} \label{eq:bk}
b_k = \int\limits_{\Omega} |V_k^{(0)}|^2 = \lim\limits_{p\to 0} \partial_p \mu_k^{(p)} 
\end{equation}
(see \cite{Friedlander91} and Appendix \ref{sec:small-p}).  In
particular, the smallest eigenvalue $\mu_0^{(p)}$ vanishes as
$\mu_0^{(p)} \approx p |\Omega|/|\pa|$, where $|\Omega|$ and $|\pa|$
are the volume of $\Omega$ and the surface area of $\pa$,
respectively, and $f\approx g$ means that the ratio $f/g$ tends to $1$
in the considered limit.  What is the small-$p$ asymptotic behavior of
$\mu_k^{(p)}$ for exterior domains (for which $|\Omega| = \infty$)?

\section{Example of a ball}
\label{sec:ball}

To gain some intuition, it is instructive to start by considering the
exterior of a ball $B_L \subset \R^d$ of radius $L$:
\begin{equation}
E_L = \{ \x\in\R^d ~:~ |\x| > L\} = \R^d \backslash \overline{B_L} ,
\end{equation}
where overline denotes the closure of $B_L$.  The rotational symmetry
of this domain and the consequent separation of variables in the
modified Helmholtz equation written in spherical coordinates
$(r,\theta,\cdots)$ allow one to find the Steklov eigenvalues and
eigenfunctions explicitly:
\begin{equation}  \label{eq:mun_ball_Rd}
\mu_n^{(p,L)} = - \alpha \frac{k'_{n,d}(\alpha L)}{k_{n,d}(\alpha L)} \,, \qquad V_{n,m}^{(p,L)}(\x) = \psi_{n,m}(\theta,\cdots) \, g_n^{(p,L)}(r)   
\qquad (n = 0,1,2,\ldots),
\end{equation}
where $\alpha = \sqrt{p}$, $r = |\x|$ is the radial coordinate, prime
denotes the derivative with respect to the argument, $k_{n,d}(z) =
z^{1-d/2} K_{n-1+d/2}(z)$ is an extension of the modified Bessel
function $K_n(z)$ of the second kind, $\{\psi_{n,m}\}$ are normalized
spherical harmonics on the sphere $\partial B_L$ in $\R^d$ (such that
$\int\nolimits_{\partial B_L} |\psi_{n,m}|^2 = 1$), and
\begin{equation}  \label{eq:gn_radial}
g_n^{(p,L)}(r) = \frac{k_{n,d}(\alpha r)}{k_{n,d}(\alpha L)}
\end{equation}
are the radial functions.  In contrast to the conventional enumeration
of Steklov eigenfunctions in Eq. (\ref{eq:Steklov}) by a single index
$k$, we employ here a multi-index $n,m$ inherited from enumeration of
spherical harmonics.  For instance, the usual spherical harmonics
$\psi_{n,m}(\theta,\phi)$ in three dimensions are enumerated by $n =
0,1,2,\ldots$ and $m= -n,-n+1,\ldots,n$.  In two dimensions, one deals
with Fourier harmonics $e^{in\theta}$ and $e^{-in\theta}$, which can
alternatively be enumerated by $n = 0,1,2,\ldots$ and $m = \pm 1$
distinguishing the sign.  In higher dimensions, $m$ is a multi-index.
As the eigenvalues $\mu_n^{(p,L)}$ do not depend on $m$, they are
generally degenerate (e.g., $\mu_n^{(p,L)}$ is $(2n+1)$ times
degenerate in three dimensions).

Setting $r = L$, one sees that spherical harmonics $\psi_{n,m}$ are
the eigenfunctions of the Dirichlet-to-Neumann operator $\M_p^L$ on
$\partial B_L$, associated with the eigenvalues $\mu_n^{(p,L)}$.  In
particular, these eigenfunctions do not depend on $p$; moreover, the
principal eigenfunction, associated to the smallest eigenvalue
$\mu_0^{(p,L)}$, is constant: $v_{0,0}^{(p,L)} = \psi_{0,0} =
1/\sqrt{|\partial B_L|}$.  

One can easily check that the functions $k_{n,d}(z)$ satisfy the
second-order differential equation
\begin{equation}  \label{eq:knd_Bessel}
k_{n,d}''(z) + \frac{d-1}{z} k_{n,d}'(z) - \biggl(1 + \frac{n(n+d-2)}{z^2}\biggr) k_{n,d}(z) = 0,
\end{equation}
and the recurrence relation 
\begin{equation}  \label{eq:knd_rec}
k'_{n,d}(z) = -k_{n-1,d}(z) - \frac{n+d-2}{z} k_{n,d}(z),
\end{equation}
that yields
\begin{equation}  \label{eq:mun_ball_Rd2}
\mu_n^{(p,L)} = \frac{n+d-2}{L} + \alpha \frac{K_{n-2+d/2}(\alpha L)}{K_{n-1+d/2}(\alpha L)} \geq 0.
\end{equation}
Since $K_\nu(z)$ is a monotonously increasing function of $\nu$ for $z
> 0$ and $\nu > 0$ (this property follows from an integral
representation of $K_\nu(z)$, see 9.6.24 from
\cite{Abramowitz}), we conclude that
\begin{equation}
0 \leq \mu_n^{(p,L)} - \mu_n^{(0,L)} \leq \sqrt{p}
\end{equation}
for any $n$ and $L$ in dimensions $d \geq 3$.

The exponential decay of the radial functions $g_n^{(p,L)}(r)$ as
$r\to \infty$, inherited from the asymptotic behavior of $K_n(z)$ at
large $z$, ensures the convergence of the $L^2(E_L)$-norm of all
Steklov eigenfunctions $V_n^{(p,L)}$ for any $p > 0$, as expected
\cite{Auchmuty13}.  In Appendix \ref{sec:integral_ball}, we compute
explicitly this norm and show that
\begin{equation}  \label{eq:intV2_dmu_ball}
\int\limits_{E_L} |V_{n,m}^{(p,L)}|^2 = \partial_p \mu_n^{(p,L)} .
\end{equation}

In the limit $p\to 0$, the asymptotic behavior of $K_n(z)$ implies
\begin{equation}  \label{eq:mun_ball_Rd_p0}
\mu_n^{(0,L)} = \frac{n+d-2}{L} \,, \qquad V_{n,m}^{(0,L)} = \psi_{n,m} \, (L/r)^{n+d-2}   
\qquad (n = 0,1,2,\ldots).
\end{equation}
The small-$p$ asymptotic behavior of the norm
$\|V_{n,m}^{(p,L)}\|_{L^2(E_L)}$ and of the associated eigenvalue
$\mu_n^{(p,L)}$ turns out to be tightly related: if the
$L^2(E_L)$-norm of $V_{n,m}^{(0,L)}$ is finite, one retrieves the
asymptotic relation (\ref{eq:muk_small_p_bounded}), derived for
bounded domains; in turn, if the norm is infinite, a non-analytic term
emerges in the leading order in $p$, as detailed below.

(i) In two dimensions, the limit of the principal eigenvalue is zero,
$\mu_0^{(0,L)} = 0$, whereas the associated Steklov eigenfunction
$V_{0,0}^{(p,L)}$ approaches a constant, $V_{0,0}^{(0,L)} =
1/\sqrt{2\pi L}$, which is not square integrable in the exterior of
the disk.  Similarly, the $L^2(E_L)$-norm of $V_{1,m}^{(0,L)}$ is also
infinite.  Accordingly, Eq. (\ref{eq:mun_ball_Rd2}) implies an
non-analytic behavior of the associated eigenvalues in the leading
order (see also \cite{Grebenkov19c}):
\begin{subequations}  \label{eq:mun_ball_2d}
\begin{align}  \label{eq:mun_ball_2d_0}
\mu_0^{(p,L)} &\approx \frac{-1}{L(\ln(\sqrt{p}L/2) + \gamma)} + O(p)\,, \\    \label{eq:mun_ball_2d_1}
\mu_1^{(p,L)} &\approx \frac{1}{L} - L p \bigl(\gamma + \ln(\sqrt{p}L/2)\bigr) ,
\end{align}
\end{subequations}
with $\gamma \approx 0.5772$ being the Euler's constant.  In turn, the
other eigenvalues exhibit a linear dependence on $p$ in the leading
order:
\begin{equation}
\mu_n^{(p,L)} \approx \frac{n}{L} + b_n^L\, p + o(p)  \qquad (n \geq 2),
\end{equation}
where
\begin{equation}  \label{eq:bnL}
b_n^L = \int\limits_{E_L} |V_{n,m}^{(0,L)}|^2 = \frac{L}{2n+d-4} 
\end{equation}
(see Appendix \ref{sec:integral_ball} for the computation of this
integral).

(ii) In three dimensions, Eq. (\ref{eq:mun_ball_Rd2}) yields
\begin{equation}  \label{eq:mun_ball_3d}
\mu_0^{(p,L)} = \frac{1}{L} + \sqrt{p} ,   \qquad  \mu_n^{(p,L)} \approx \frac{n+1}{L} + b_n^L\, p  + o(p)  \quad (n \geq 1),
\end{equation}
where the first relation for $\mu_0^{(p,L)}$ is exact and valid for
any $p\geq 0$.  In other words, the asymptotic relation
(\ref{eq:muk_small_p_bounded}) fails for the principal eigenvalue, in
agreement with the infinite $L^2(E_L)$-norm of $V_{0,0}^{(0,L)} =
1/(\sqrt{4\pi} ~|\x|)$.

(iii) In four dimensions, the integral in Eq. (\ref{eq:bnL}) is
infinite for $n = 0$, whereas the smallest eigenvalue $\mu_0^{(p,L)}$
behaves as
\begin{equation}  \label{eq:mun_ball_4d}
\mu_0^{(p,L)} \approx \frac{2}{L} - Lp \bigl(\gamma + \ln(\sqrt{p}L/2)\bigr) + O(p).  
\end{equation}
In turn, the other coefficients $b_n^L$ are finite and one retrieves
the linear asymptotic behavior (\ref{eq:muk_small_p_bounded}) for
other eigenvalues.  

(iv) Finally, for $d > 4$, the integral in Eq. (\ref{eq:bnL}) is
finite for all eigenfunctions and Eq. (\ref{eq:muk_small_p_bounded})
holds for all eigenvalues $\mu_n^{(p,L)}$.  For instance, one has
$\mu_0^{(p,L)} = 3/L + p L/(1 + \sqrt{p} L) \approx 3/L + p L +
O(p^{3/2})$ for $d = 5$, i.e., the leading term is linear in $p$,
whereas the non-analytic term appears in the next order.

In the next section, we extend the above results to other exterior
domains and uncover the asymptotic behavior of the eigenvalues as
$p\to 0$.

\section{Asymptotic behavior of eigenvalues}
\label{sec:asympt}

Let $V_k^{(p)}$ and $\mu_k^{(p)}$ be the Steklov eigenfunctions and
eigenvalues in the exterior domain $\Omega = \R^d \backslash
\Omega_0$, where $\Omega_0$ is a compact set in $\R^d$.  We aim
at deriving the asymptotic behavior of $\mu_k^{(p)}$ as $p\to 0$.  For
this purpose, we split $\Omega$ into a bounded domain $\Omega_L =
\Omega \cap B_L$ and the exterior $E_L$ of a ball $B_L$ of radius $L$
large enough such that $\Omega_0 \subset B_L$.
We start by deducing three identities.

\subsection{Three identities}
\label{sec:identities}

First, we integrate the equation $(p-\Delta)V_k^{(p)} = 0$ over the
domain $\Omega_L$ to get
\begin{equation}
0 = p \int\limits_{\Omega_L} V_k^{(p)} - \int\limits_{\pa \cup \partial B_L} \partial_n V_k^{(p)} ,
\end{equation}
so that
\begin{equation}  \label{eq:auxil13}
p \int\limits_{\Omega_L} V_k^{(p)} = \mu_k^{(p)} \int\limits_{\pa} v_k^{(p)} + \int\limits_{\partial B_L} \partial_n V_k^{(p)} ,
\end{equation}
where we used the Green's formula.  As $V_k^{(p)}$ is an analytic
function away from the boundary $\pa$, its normal derivative on
$\partial B_L$ should be continuous:
\begin{equation}
(\partial_n V_k^{(p)})|_{\partial B_L^{-}} = - (\partial_n V_k^{(p)})|_{\partial B_L^{+}} ,
\end{equation}
where $\partial B_L^{-} = \partial \Omega_L$ and $\partial B_L^{+} =
\partial E_L$ denote the spherical boundary $\partial B_L$ approached
from inside and from outside, respectively, and the minus sign emerges
from the opposite directions of the normal derivative on two sides of
that boundary.  The right-hand side of the above relation can then be
expressed with the help of an auxiliary Dirichlet-to-Neumann operator
$\M_p^L$ in the exterior of the ball $B_L$, implying
\begin{equation}  \label{eq:BC_MpL}
\partial_n V_k^{(p)} = - \M_p^L V_k^{(p)}  \quad \textrm{on}~\partial B_L
\end{equation}
(here we dropped the superscript minus for brevity).  Integrating this
expression over $\partial B_L$ and interpreting the integral as the
scalar product in $L^2(\partial B_L)$, one gets
\begin{equation}
\int\limits_{\partial B_L} \partial_n V_k^{(p)} = - \bigl(\M_p^L V_k^{(p)}, 1\bigr)_{L^2(\partial B_L)}
= - \bigl( V_k^{(p)}, \M_p^L 1\bigr)_{L^2(\partial B_L)} = - \mu_0^{(p,L)}  \bigl( V_k^{(p)}, 1\bigr)_{L^2(\partial B_L)}
= - \mu_0^{(p,L)} \int\limits_{\partial B_L} V_k^{(p)} ,
\end{equation}
where $\mu_0^{(p,L)}$ is the principal eigenvalue of $\M_p^L$,
associated to the constant eigenfunction (on the sphere $\partial
B_L$).  Substituting this expression into Eq. (\ref{eq:auxil13}), we
deduce the first identity
\begin{equation}  \label{eq:identity1}
\mu_k^{(p)} \int\limits_{\pa} v_k^{(p)} = p \int\limits_{\Omega_L} V_k^{(p)} + \mu_0^{(p,L)}  \int\limits_{\partial B_L} V_k^{(p)} .
\end{equation}
In the limit $p\to 0$, one gets
\begin{equation}  \label{eq:identity1_p0}
\mu_k^{(0)} \int\limits_{\pa} v_k^{(0)} = \frac{d-2}{L} \int\limits_{\partial B_L} V_k^{(0)} .
\end{equation}

Second, we multiply $(p - \Delta)V_k^{(p)} = 0$ by $[V_k^{(p)}]^*$ and
integrate over $\Omega_L$ to get
\begin{equation*}
0 = \int\limits_{\Omega_L} \biggl(p |V_k^{(p)}|^2 + |\nabla V_k^{(p)}|^2\biggr) - \mu_k^{(p)} \underbrace{\int\limits_{\pa} |v_k^{(p)}|^2}_{=1}
- \int\limits_{\partial B_L} [V_k^{(p)}]^* \partial_n V_k^{(p)} .
\end{equation*}
Using Eq. (\ref{eq:BC_MpL}) and
the eigenbasis of the operator $\M_p^L$, one obtains the following
representation for the eigenvalue:
\begin{equation}   \label{eq:muk_int}
\mu_k^{(p)} = \int\limits_{\Omega_L} \biggl(p |V_k^{(p)}|^2 + |\nabla V_k^{(p)}|^2\biggr)
+ \sum\limits_{n,m} \mu_n^{(p,L)} \bigl|\bigl(V_k^{(p)}, \psi_{n,m}\bigr)_{L^2(\partial B_L)}\bigr|^2 .
\end{equation}
In particular, one has in the limit $p\to 0$:
\begin{equation}   \label{eq:muk_int_p0}
\mu_k^{(0)} = \int\limits_{\Omega_L} |\nabla V_k^{(0)}|^2
+ \sum\limits_{n,m} \mu_n^{(0,L)} \bigl|\bigl(V_k^{(0)}, \psi_{n,m}\bigr)_{L^2(\partial B_L)}\bigr|^2 .
\end{equation}
This identity implies that all eigenvalues $\mu_k^{(0)}$ are strictly
positive for $d \geq 3$: 
\begin{equation}
\mu_k^{(0)} > 0  \qquad (k = 0,1,2,\ldots).
\end{equation}
Indeed, as $\mu_n^{(0,L)} = (n + d-2)/L > 0$ for all $n$ and $d \geq
3$, an eigenvalue $\mu_k^{(0)}$ could be zero only if all terms in
Eq. (\ref{eq:muk_int_p0}) are strictly zero, i.e., if
$V_k^{(0)}|_{\partial B_L}$ was orthogonal to {\it all} spherical
harmonics $\psi_{n,m}$ that would contradict the completeness of
$\{\psi_{n,m}\}$ in $L^2(\partial B_L)$.  In other words, the
condition $\mu_k^{(0)} = 0$ would require that $V_k^{(0)}|_{\partial
B_L} \equiv 0$ for any $L$ large enough, that is impossible.  Note
that this argument is not applicable in two dimensions because
$\mu_0^{(0,L)} = 0$, see Sec. \ref{sec:2D_asympt} below.

Third, multiplying $(p-\Delta) V_k^{(p)} = 0$ by $[V_k^{(p')}]^*$,
multiplying $(p'-\Delta) [V_k^{(p')}]^* = 0$ by $V_k^{(p)}$,
subtracting these equations and integrating over $\Omega_L$, we have
\begin{align*} \nonumber
0 & = (p - p') \int\limits_{\Omega_L} V_k^{(p)} [V_k^{(p')}]^* 
- \int\limits_{\Omega_L} \bigl([V_k^{(p')}]^* \Delta V_k^{(p)} - V_k^{(p)} \Delta [V_k^{(p')}]^*\bigr) \\
\nonumber
& = (p - p') \int\limits_{\Omega_L} V_k^{(p)} [V_k^{(p')}]^* 
- \int\limits_{\pa \cup \partial B_L} \bigl([V_k^{(p')}]^* \partial_n V_k^{(p)} - V_k^{(p)} \partial_n [V_k^{(p')}]^*\bigr) \\  
& = (p - p') \int\limits_{\Omega_L} V_k^{(p)} [V_k^{(p')}]^* - [\mu_k^{(p)} - \mu_k^{(p')}] \int\limits_{\pa} v_k^{(p)} [v_k^{(p')}]^* 
+ \int\limits_{\partial B_L} \biggl([V_k^{(p')}]^* \M_p^L V_k^{(p)} - V_k^{(p)} \M_{p'}^L [V_k^{(p')}]^*\biggr).
\end{align*}
We get thus the third identity:
\begin{equation}  \label{eq:identity2p}
[\mu_k^{(p)} - \mu_k^{(p')}] \int\limits_{\pa} v_k^{(p)} [v_k^{(p')}]^*  = (p - p') \int\limits_{\Omega_L} V_k^{(p)} [V_k^{(p')}]^* 
+ W_k^{(p,p',L)} ,
\end{equation}
where
\begin{align}  \nonumber
W_k^{(p,p',L)} & = \int\limits_{\partial B_L} \biggl([V_k^{(p')}]^* \M_p^L V_k^{(p)} - V_k^{(p)} \M_{p'}^L [V_k^{(p')}]^*\biggr)
= \bigl(V_k^{(p')}, (\M_p^L - \M_{p'}^L) V_k^{(p)}\bigr)_{L^2(\partial B_L)}  \\  \label{eq:Wk}
& = \sum\limits_{n,m} \bigl(V_k^{(p')} ,\psi_{n,m}\bigr)_{L^2(\partial B_L)}
\bigl[\mu_n^{(p,L)} - \mu_n^{(p',L)}\bigr] \bigl(\psi_{n,m}, V_k^{(p)}\bigr)_{L^2(\partial B_L)} .
\end{align}
Here we used that the Dirichlet-to-Neumann operator $\M_{p'}^L$ is
self-adjoint, while its eigenfunctions $\{ \psi_{n,m}\}$ do not depend
on $p$ and $p'$ (see Sec. \ref{sec:ball}).  Note that the scalar
products in Eq. (\ref{eq:Wk}) can be expressed by using the identity
(\ref{eq:fourth_identity}) from Appendix \ref{sec:proof}.

The third identity has two important consequences.  (i) Setting $p' =
0$, we get another identity that will be employed in the following
analysis:
\begin{equation}  \label{eq:identity2}
[\mu_k^{(p)} - \mu_k^{(0)}] \int\limits_{\pa} [v_k^{(0)}]^* v_k^{(p)}  = p \int\limits_{\Omega_L} [V_k^{(0)}]^* V_k^{(p)} 
+ W_k^{(p,0,L)} .
\end{equation}
(ii) Since the generalized exterior Steklov problem is well posed for
$p > 0$, one can set $p' = p + \epsilon$ and use a smooth dependence
of $\mu_k^{(p)}$ and $V_k^{(p)}$ on $p$ and boundness of $\Omega_L$ to
develop Eq. (\ref{eq:identity2p}) in the limit $\epsilon\to 0$
\begin{equation} 
\partial_p \mu_k^{(p)} \underbrace{\int\limits_{\pa} |v_k^{(p)}|^2}_{=1} = 
\int\limits_{\Omega_L} |V_k^{(p)}|^2 +
\sum\limits_{n,m} \bigl|\bigl(V_k^{(p)} ,\psi_{n,m}\bigr)_{L^2(\partial B_L)}\bigr|^2 \partial_p \mu_n^{(p,L)}  .
\end{equation}
Substituting Eq. (\ref{eq:intV2_dmu_ball}), one sees that the second
term in the right-hand side is the integral of $|V_k^{(p)}|^2$ over
the exterior $E_L$.  Indeed, one has
\begin{align*}
\int\limits_{E_L} |V_k^{(p)}|^2 & = \sum\limits_{n,m,n'm'} \bigl(V_k^{(p)} ,\psi_{n,m}\bigr)_{L^2(\partial B_L)}
\bigl(\psi_{n',m'}, V_k^{(p)}\bigr)_{L^2(\partial B_L)}  \int\limits_{E_L} [V_{n,m}^{(p,L)}]^* V_{n',m'}^{(p,L)} \\
& = \sum\limits_{n,m} \bigl|\bigl(V_k^{(p)} ,\psi_{n,m}\bigr)_{L^2(\partial B_L)}\bigr|^2 
\underbrace{\int\limits_{E_L} |V_{n,m}^{(p,L)}|^2}_{= \partial_p \mu_n^{(p,L)}} ,
\end{align*}
where the integral over angular coordinates canceled all the terms
except if $n = n'$ and $m = m'$.  As a consequence, we obtain
\begin{equation}  \label{eq:dmu}
\partial_p \mu_k^{(p)} = \int\limits_{\Omega} |V_k^{(p)}|^2  \qquad (p > 0).
\end{equation}
This relation, which was earlier derived for bounded domains
\cite{Friedlander91}, allows us to distinguish two scenarios in the
limit $p\to 0$: if the squared $L^2(\Omega)$-norm of the limiting
eigenfunction $V_k^{(0)}$ is finite, the associated eigenvalue
$\mu_k^{(p)}$ approaches its limit $\mu_k^{(0)}$ {\it linearly} with
$p$, as in Eq. (\ref{eq:muk_small_p_bounded}) for bounded domains.  In
turn, if this norm is infinite, the derivative $\partial_p
\mu_k^{(p)}$ is also infinite, suggesting an non-analytic behavior of
$\mu_k^{(p)}$ near $p = 0$.

In the next subsection, we employ these identities to derive the
asymptotic behavior of $\mu_k^{(p)}$ as $p\to 0$.

\subsection{Two-dimensional case}
\label{sec:2D_asympt}

In two dimensions, the identity (\ref{eq:identity1_p0}) implies for
any $k$ that either $\mu_k^{(0)} = 0$, or the integral of $v_k^{(0)}$
over the boundary $\pa$ is zero.  Christiansen and Datchev showed that
$\mu_0^{(0)} = 0$, while $v_0^{(0)}$ is a constant function
\cite{Christiansen23}.  According to Eq. (\ref{eq:muk_int_p0}),
$\mu_0^{(0)} = 0$ implies that $V_0^{(0)}$ is also a constant
function.  Since such a constant function is unique (up to a
multiplicative factor), the eigenvalue $\mu_0^{(0)}$ is simple, i.e.,
the next eigenvalue is strictly positive: $\mu_1^{(0)} > 0$.  We
conclude that
\begin{align}   \label{eq:v0-2d}
\mu_0^{(0)} & = 0,  \qquad v_0^{(0)} = \frac{1}{\sqrt{|\pa|}} \,, \\  \label{eq:v0-2d_k}
\mu_k^{(0)} & > 0 , \qquad  \int\limits_{\pa} v_k^{(0)} = 0  \qquad (k > 0).
\end{align}  

The identity (\ref{eq:identity2}) yields in the leading order:
\begin{equation}
\mu_k^{(p)} - \mu_k^{(0)} = \frac{\mu_0^{(p,L)}}{2\pi L} \biggl|\int\limits_{\partial B_L} V_k^{(p)}\biggr|^2 + O(1/\ln^2(\sqrt{p})),
\end{equation}
where we wrote explicitly the leading-order term of $W_k^{(p,0,L)}$
from Eq. (\ref{eq:Wk}) with $n = 0$ (for which $\psi_{0,0} =
1/\sqrt{2\pi L}$), used the normalization of $v_k^{(0)}$ to replace
$\int\limits_\pa [v_k^{(0)}]^* v_k^{(p)}$ by $1$ in the leading order,
and ignored higher-order terms.   Using Eq. (\ref{eq:mun_ball_2d_0}),
we get in the leading order
\begin{equation}  \label{eq:muk_2D_conjectural}  
\mu_k^{(p)} - \mu_k^{(0)} = \frac{-a_k}{\ln(\sqrt{p})} + O(1/\ln^2(\sqrt{p})),
\end{equation}
where
\begin{equation}  \label{eq:ck_def}  
a_k = 2\pi |c_k|^2, \qquad 
c_k = \frac{1}{2\pi  L} \int\limits_{\partial B_L} V_k^{(0)} 
= \int\limits_0^{2\pi} \frac{d\theta}{2\pi} \, V_k^{(0)}(L,\theta)
\end{equation}
(in the last equality, we used polar coordinates $(r,\theta)$ with the
origin at the center of the disk $B_L$).
It is important to stress that the coefficient $c_k$ does not depend
on the choice of $L$.  Indeed, the Steklov eigenfunction outside the
disk of radius $L$ reads in polar coordinates $(r,\theta)$ as
\begin{equation}  \label{eq:Vk0_outside_disk}
V_k^{(0)}(r,\theta) = \sum\limits_{n=-\infty}^\infty \biggl(\frac{L}{r}\biggr)^{|n|} 
e^{in\theta} \int\limits_0^{2\pi} \frac{d\theta'}{2\pi} \, e^{-in\theta'} V_k^{(0)}(L,\theta')  \qquad (r \geq L).
\end{equation}
One sees that the term with $n = 0$ is precisely the coefficient
$c_k$, which therefore determines a constant term in the asymptotic
behavior of $V_k^{(0)}$ at large $|\x|$:
\begin{equation}  \label{eq:Vk_inf}
V_k^{(0)}(\x) \simeq c_k + O(1/|\x|) \qquad |\x|\to \infty.
\end{equation}
When $c_k \ne 0$, the eigenvalue $\mu_k^{(p)}$ exhibits the
logarithmically slow approach (\ref{eq:muk_2D_conjectural}) to its
limit $\mu_k^{(0)}$, while the associated eigenfunction approaches
$c_k$ at infinity.

For instance, this asymptotic behavior holds for the principal
eigenvalue with $k = 0$.  Indeed, as $v_0^{(0)} = 1/\sqrt{|\pa|}$, the
associated eigenfunction $V_0^{(0)}$ is also constant, implying $c_0 =
1/\sqrt{|\pa|}$, and one gets $a_k = 2\pi/|\pa|$ and thus
\begin{equation}  \label{eq:auxil12}
\mu_0^{(p)} \approx \frac{-2\pi}{|\pa| \ln(\sqrt{p})} + O(1/\ln^2(\sqrt{p})) \quad (p\to 0).
\end{equation}
Note that next-order terms determine the correct lengthscale that is
needed to make the argument of the logarithm in the denominator
dimensionless.  Christiansen and Datchev analyzed the small-$p$
asymptotic behavior for connected unbounded domains $\Omega = \R^2
\backslash \Omega_0$ with a smooth boundary $\pa$ and identified this
lengthscale as the logarithmic capacity $R_c$ of the compact set
$\Omega_0$.  In fact, they proved \cite{Christiansen23}
\begin{equation}\label{eq:mu0-2d}
\mu_0^{(p)} \approx \frac{- 2 \pi}{|\pa| (\ln (R_c \sqrt{p} / 2) + \gamma)} + 
O\left(1/\ln^2(\sqrt{p})\right) \quad (p \rightarrow 0).
\end{equation}
Even though this rigorous result has the same leading-order behavior
as Eq. (\ref{eq:auxil12}), it turns out to be more accurate, as will
be checked by numerical simulations in Sec. \ref{sec:2D}.  Moreover,
it provides a proper length $R_c$ to get a dimensionless argument of
the logarithm.

For $k > 0$, we apply the identity (\ref{eq:identity1}) in the leading
order,
\begin{equation}  
\mu_k^{(p)} \int\limits_{\pa} v_k^{(p)} = \mu_0^{(p,L)}  \int\limits_{\partial B_L} V_k^{(p)} + O(p),
\end{equation}
as well as Eq. (\ref{eq:mun_ball_2d_0}, \ref{eq:muk_2D_conjectural}),
to get
\begin{equation}  
\mu_k^{(0)} \int\limits_{\pa} v_k^{(p)}  
= \frac{-1}{\ln(\sqrt{p})} \biggl(\frac{1}{L}\int\limits_{\partial B_L} V_k^{(0)}\biggr) + \ldots,
\end{equation}
where $\ldots$ represents the next-order term.  Using
Eqs. (\ref{eq:v0-2d_k}, \ref{eq:ck_def}), we have
\begin{equation}  \label{eq:vk_int_2d}
\int\limits_{\pa} \bigl(v_k^{(p)} - v_k^{(0)}\bigr) = -\frac{2\pi c_k}{\mu_k^{(0)} \ln(\sqrt{p})} + O(1/\ln^2(\sqrt{p})) 
\qquad (k > 0)\,.
\end{equation}
As a consequence, if $c_k \ne 0$, the eigenfunction $v_k^{(p)}$
approaches its limit $v_k^{(0)}$ logarithmically slowly,
\begin{equation}  \label{eq:vk_log_2d}
v_k^{(p)} = v_k^{(0)} + \frac{u_k}{\ln(\sqrt{p})} + O(1/\ln^2(\sqrt{p}))  \qquad (k > 0),
\end{equation}
where the leading-order correction term $u_k$ determines
the coefficients $a_k$ and $c_k$:
\begin{equation}   \label{eq:ak_2D}
c_k = - \frac{\mu_k^{(0)}}{2\pi} \int\limits_{\pa} u_k \,, \qquad
a_k = \frac{[\mu_k^{(0)}]^2}{2\pi}  \biggl| \int\limits_{\pa} u_k\biggr|^2 \qquad (k > 0).
\end{equation}
This expression relates the asymptotic behavior of $V_k^{(0)}$ at
infinity to the asymptotic behavior of $v_k^{(p)}$ on the boundary
$\pa$.

What does happen in the case $c_k = 0$?  The asymptotic behavior of
$\mu_k^{(p)}$ is determined by the next-order term, which comes from
the term $n = 1$ in Eq. (\ref{eq:Wk}).  Indeed, the identity
(\ref{eq:identity2}) yields in the leading order:
\begin{equation}
\mu_k^{(p)} - \mu_k^{(0)} = \frac{\mu_1^{(p,L)}}{2\pi L} \biggl\{
\biggl|\int\limits_{\partial B_L} V_k^{(0)} e^{i\theta} \biggr|^2 +
\biggl|\int\limits_{\partial B_L} V_k^{(0)} e^{-i\theta} \biggr|^2\biggr\} + O(p),
\end{equation}
where we wrote explicitly the Fourier harmonics $\psi_{1,m} =
e^{im\theta}/\sqrt{2\pi L}$ in polar coordinates (with $m = \pm 1$).
Substituting the leading-order term in the asymptotic relation
(\ref{eq:mun_ball_2d_1}) for $\mu_1^{(p,L)}$, we find
\begin{equation}  \label{eq:muk_2d_plogp}
\mu_k^{(p)} - \mu_k^{(0)} = -p\ln(\sqrt{p}) \, d_k + O(p),  
\end{equation}
where
\begin{equation}  \label{eq:dk_def}
d_k = \frac{L^2}{2\pi} \biggl\{
\biggl|\int\limits_0^{2\pi} d\theta\, V_k^{(0)}(L,\theta) e^{i\theta} \biggr|^2 +
\biggl|\int\limits_0^{2\pi} d\theta\, V_k^{(0)}(L,\theta) e^{-i\theta} \biggr|^2\biggr\} .
\end{equation}
Using the identity (\ref{eq:fourth_identity}) derived in Appendix
\ref{sec:proof}, we can express the coefficient $d_k$ through the
integral over the boundary $\pa$.  In fact,
Eq. (\ref{eq:fourth_identity}) reads in two dimensions for $n = 1$ and
$m = \pm 1$:
\begin{equation}
L \int\limits_0^{2\pi} d\theta \, V_k^{(0)}(L,\theta) \, e^{im\theta} = \frac12 \int\limits_{\pa} v_k^{(0)}
\biggl(\mu_k^{(0)} r e^{im\theta} - \partial_n (r e^{im\theta})\biggr),
\end{equation}
where the polar coordinates $(r,\theta)$ were also used inside the
disk $B_L$.  We stress that both $r$ and $\theta$ in the right-hand
side depend on the boundary point.  As a consequence, the coefficient
$d_k$ does not depend on $L$.  While the coefficient $c_k$ determined
the constant term in the asymptotic behavior of $V_k^{(0)}$ at
infinity, the coefficient $d_k$ controls the $O(1/|\x|)$ term.

Finally, if both $c_k$ and $d_k$ are zero, then $\mu_k^{(p)} -
\mu_k^{(0)}$ in the identity (\ref{eq:identity2}) is controlled
by $O(p)$ terms.  In order to determine this asymptotic behavior, we
note that both $O(1)$ and $O(1/|\x|)$ terms in
Eq. (\ref{eq:Vk0_outside_disk}) are cancelled, i.e., $V_k^{(0)}$
behaves as $O(1/|\x|^2)$ at infinity, so that its $L^2(\Omega)$ is
finite.  As a consequence, the limit of Eq. (\ref{eq:dmu}) as $p\to 0$
determines the derivative $\partial_p \mu_k^{(p)}$ at $p = 0$, and
thus the asymptotic relation (\ref{eq:muk_small_p_bounded}) holds,
with $b_k$ given by Eq. (\ref{eq:bk}), as for bounded domains.

The above findings can be summarized as follows:

(i) If $c_k \ne 0$, then $V_k^{(0)}(\x) \simeq c_k$ as $|\x|\to
\infty$, and the asymptotic relation (\ref{eq:muk_2D_conjectural})
holds;

(ii) If $c_k = 0$ and $d_k \ne 0$, then $V_k^{(0)} \simeq O(1/|\x|)$
as $|\x|\to \infty$, the $L^2(\Omega)$-norm of $V_k^{(0)}$ is
infinite, and the asymptotic relation (\ref{eq:muk_2d_plogp}) holds;

(iii) If $c_k = d_k = 0$, then $V_k^{(0)} \simeq O(1/|\x|^\nu)$ as
$|\x|\to\infty$ with $\nu \geq 2$, the $L^2(\Omega)$-norm of
$V_k^{(0)}$ is finite, and the asymptotic relation
(\ref{eq:muk_small_p_bounded}) holds.

\subsection{Three-dimensional case}

In three dimensions, one can substitute the asymptotic relations
(\ref{eq:mun_ball_3d}) into Eq. (\ref{eq:Wk}) to get in the leading
order:
\begin{equation}
W_k^{(p,0,L)} = \sqrt{p}\, \bigl|\bigl(V_k^{(0)} ,\psi_{0,0}\bigr)_{L^2(\partial B_L)}\bigr|^2 + O(p) ,
\end{equation}
where $\psi_{0,0} = 1/\sqrt{4\pi L^2}$ is the constant eigenfunction
of $\M_p^L$ on $\partial B_L$.  As a consequence, the identity
(\ref{eq:identity2}) implies in the leading order:
\begin{equation}  
[\mu_k^{(p)} - \mu_k^{(0)}] \underbrace{\int\limits_{\pa} |v_k^{(0)}|^2}_{=1} = 
\frac{\sqrt{p}}{4\pi L^2} \biggl|\int\limits_{\partial B_L} V_k^{(0)}\biggr|^2 + O(p). 
\end{equation}
Using the identity (\ref{eq:identity1_p0}), we find
\begin{equation}    \label{eq:muk_ak_3d}
\mu_k^{(p)} = \mu_k^{(0)} + a_k \sqrt{p} + O(p),
\end{equation}
with
\begin{equation}   \label{eq:ak}
a_k = \frac{[\mu_k^{(0)}]^2}{4\pi} \biggl|\int\limits_{\pa} v_k^{(0)}\biggr|^2 .
\end{equation}
One sees that the integral of $v_k^{(0)}$ over the boundary $\pa$
determines the small-$p$ asymptotic behavior of the associated
eigenvalue $\mu_k^{(p)}$.  Note that Cauchy-Schwarz inequality implies
an upper bound for the coefficients $a_k$:
\begin{equation}
a_k \leq \frac{[\mu_k^{(0)}]^2}{4\pi} |\pa| .
\end{equation}

When the integral of $v_k^{(0)}$ over $\pa$ is zero, i.e., $a_k = 0$,
one needs to search for the next-order term, which is linear in $p$.
Substituting Eq. (\ref{eq:mun_ball_3d}) for $n > 0$ into
Eq. (\ref{eq:Wk}), we get
\begin{equation}  \label{eq:Wk_p1a}
W_k^{(p,0,L)} = p \sum\limits_{n,m} \bigl|\bigl(V_k^{(0)} ,\psi_{n,m}\bigr)_{L^2(\partial B_L)}\bigr|^2
\int\limits_{E_L} |V_{n,m}^{(0,L)}|^2 + o(p)  ,
\end{equation}
where there is no term with $n = 0$ (that yielded the $\sqrt{p}$
contribution in the above analysis) because $\bigl(V_k^{(0)}
,\psi_{0,0}\bigr)_{L^2(\partial B_L)} = 0$.  Using Eqs. (\ref{eq:Qn0},
\ref{eq:u2_int}), one can evaluate the sum in Eq. (\ref{eq:Wk_p1a}) as
\begin{equation} \label{eq:Wk_p1}
W_k^{(p,0,L)} = p \int\limits_{E_L} |V_k^{(0)}|^2 + o(p).
\end{equation}
Substituting Eq. (\ref{eq:Wk_p1}) into the identity
(\ref{eq:identity2}), we get in the leading order:
\begin{equation*} 
[\mu_k^{(p)} - \mu_k^{(0)}] = p \int\limits_{\Omega_L} |V_k^{(0)}|^2
+ p \int\limits_{E_L} |V_k^{(0)}|^2 + o(p)  = p\, b_k + o(p),
\end{equation*}
where $b_k$ is the squared $L^2(\Omega)$-norm of $V_k^{(0)}$ defined
in Eq. (\ref{eq:bk}).  We therefore obtained the asymptotic behavior
(\ref{eq:muk_small_p_bounded}), which was earlier derived for bounded
domains.

We conclude that each eigenvalue $\mu_k^{(p)}$ of the generalized
exterior Steklov problem in three dimensions exhibits one of two
asymptotic relations:
\begin{align}  \label{eq:muk_3d_asympt}
\begin{cases} \textrm{(i)~ If}~~  \int\nolimits_{\pa} v_k^{(0)} \ne 0 \quad \Rightarrow \quad a_k > 0 , ~ b_k = +\infty ,
\quad \mu_k^{(p)} = \mu_k^{(0)} + a_k \sqrt{p} + O(p), \cr
\textrm{(ii)~If} ~~ \int\nolimits_{\pa} v_k^{(0)} = 0  \quad \Rightarrow \quad a_k = 0 , ~ b_k < +\infty , 
\quad \mu_k^{(p)} = \mu_k^{(0)} + b_k \,p + o(p),\end{cases} 
\end{align}
where $a_k$ and $b_k$ are determined by Eqs. (\ref{eq:ak},
\ref{eq:bk}), respectively.

For the exterior of a ball, the principal eigenfunction $v_0^{(0)}$ is
constant so that other eigenfunctions $v_k^{(0)}$ are orthogonal to
it, implying $a_k = 0$ for any $k > 0$ and thus the asymptotic
behavior (\ref{eq:muk_small_p_bounded}) for the eigenvalues
$\mu_k^{(p)}$, in agreement with the explicit result
(\ref{eq:mun_ball_3d}).  For a general exterior domain in $\R^3$,
however, there is no reason for $v_0^{(0)}$ to be constant so that
both asymptotic relations (\ref{eq:muk_small_p_bounded}) and
(\ref{eq:muk_ak_3d}) are possible.  For instance, the Steklov
eigenfunctions for the exterior of prolate and oblate spheroids were
shown to be non-constant \cite{Grebenkov24}.  In Sec. \ref{sec:3D}, we
solve numerically the generalized exterior Steklov problem for
different domains and illustrate the predicted asymptotic behavior of
$\mu_k^{(p)}$, as well as the pivotal role of the integral of
$v_k^{(0)}$ over the boundary $\pa$.

We complete this subsection by considering the asymptotic behavior of
$V_k^{(0)}$ at infinity (the following arguments are valid for $d \geq
3$).  Let us assume that
\begin{equation}  \label{eq:Vk0_asympt}
V_k^{(0)} \simeq C_k \, |\x|^{-\gamma_k}  \qquad (|\x|\to\infty), 
\end{equation}
with some constants $C_k$ and $\gamma_k$.  Substituting this behavior
into the identity (\ref{eq:identity1_p0}), one gets at large enough
$L$:
\begin{equation}  \label{eq:identity1_p0app}
\mu_k^{(0)} \int\limits_{\pa} v_k^{(0)} \approx (d-2)S_d C_k L^{d-2-\gamma_k} ,
\end{equation}
where $S_d = 2\pi^{d/2}/\Gamma(d/2)$ is the surface area of the unit
sphere in $\R^d$.  If the integral of $v_k^{(0)}$ over $\pa$ is not
zero, one has
\begin{equation}  \label{eq:gammak}
\gamma_k = 2-d,  \qquad C_k = \frac{\mu_k^{(0)}}{(d-2)S_d} \int\limits_{\pa} v_k^{(0)} .
\end{equation}
In contrast, if the integral of $v_k^{(0)}$ over $\pa$ is zero, the
relation (\ref{eq:identity1_p0app}) cannot hold for any finite $L$,
and the initial assumption (\ref{eq:Vk0_asympt}) of an isotropic
asymptotic behavior of $V_k^{(0)}$ fails.

\subsection{Higher-dimensional cases}

In four dimensions, we use the asymptotic behavior
(\ref{eq:mun_ball_4d}) of the principal eigenvalue $\mu_0^{(p,L)}$ to
get in the leading order
\begin{equation}
W_k^{(p,0,L)} \approx - Lp \ln(\sqrt{p}) \, \frac{1}{2\pi^2 L^3} \biggl|\int\limits_{\partial B_L} V_k^{(0)}\biggr|^2 + O(p),
\end{equation}
where $2\pi^2 L^3$ is the surface area of the sphere $\partial B_L$ in
$\R^4$.  From the identity (\ref{eq:identity1}), one can express the
integral of $V_k^{(0)}$ over $\partial B_L$ to get
\begin{equation}
W_k^{(p,0,L)} \approx - p \ln(\sqrt{p}) \, \frac{[\mu_k^{(0)}]^2}{8\pi^2}  \biggl|\int\limits_{\pa} v_k^{(0)}\biggr|^2 + O(p),
\end{equation}
and thus
\begin{equation}
\mu_k^{(p)} = \mu_k^{(0)} - p \ln(\sqrt{p}) \, \frac{[\mu_k^{(0)}]^2}{8\pi^2}  \biggl|\int\limits_{\pa} v_k^{(0)}\biggr|^2 + O(p).
\end{equation}
Note that the linear term $O(p)$ determines the proper length scale to
be inserted in $\ln(\sqrt{p})$, to make its argument dimensionless.
As earlier, if the integral of $v_k^{(0)}$ over $\pa$ vanishes, then
the leading term is linear in $p$.

When $d > 4$, one has $\mu_n^{(p,L)} - \mu_n^{(0,L)} \approx b_n^L \,
p + o(p)$ for all Steklov eigenvalues for the exterior of a ball, so
that the asymptotic relation (\ref{eq:Wk_p1}) holds for all $k$,
implying Eq. (\ref{eq:muk_small_p_bounded}) for all Steklov
eigenvalues $\mu_k^{(p)}$.

\section{Numerical results}
\label{sec:2D}

In this section, we undertake a numerical study of the generalized
exterior Steklov problem in two and three dimensions.  To our
knowledge, the exterior of a ball is the only exterior domain, for
which the eigenvalues and eigenfunctions of the Steklov problem are
known explicitly (see Sec. \ref{sec:ball}).  We aim therefore to
clarify how the shape of the exterior domain can affect the behavior
of the Steklov eigenvalues and eigenfunctions.  For this purpose, we
adapt a finite-element method (FEM) that was developed for interior
problems \cite{Chaigneau24} by employing the transparent boundary
condition to reduce the generalized exterior Steklov problem to an
equivalent interior one (see details and numerical validation in
Appendix \ref{sec:numerics}).

\subsection{Two dimensions}

\subsubsection*{Eigenvalues}

\begin{figure}[t!]
\includegraphics[width=\linewidth]{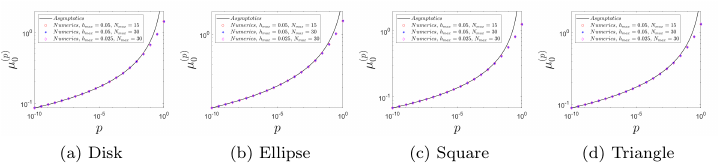}
\caption{
The small-$p$ asymptotic behavior of $\mu_0^{(p)}$ for the exterior of
{\bf (a)} a disk of radius $R=1$; {\bf (b)} an ellipse with semiaxes
$a=1$ and $b=0.5$; {\bf (c)} a square of sidelength $R=2$; and {\bf
(d)} an equilateral triangle of sidelength $R = 2$.  Solid line shows
the asymptotic relation (\ref{eq:mu0-2d}), while circles, asterisks
and diamonds present the numerical results obtained with different
mesh sizes and truncation orders (see Appendix \ref{sec:numerics} for
details), that we used to highlight the numerical accuracy.}
\label{fig:mu0-2d}
\end{figure}

First, we numerically verify the logarithmic decay (\ref{eq:mu0-2d})
of the principal eigenvalue $\mu_0^{(p)}$ for smooth shapes and simple
polygons, for which the logarithmic capacity $R_c$ is analytically
known\cite{Landkof,Ransford07,Ransford10}: (i) $R_c=R$ for a disk of
radius $R$, (ii) $R_c=(a+b)/2$ for an ellipse with semiaxes $a$ and
$b$, (iii) $R_c = \frac{\Gamma(1/4)^2}{4\pi^{3/2}} R$ for a square of
sidelength $R$, and (iv) $R_c=\frac{\Gamma(1/3)^3}{2\pi^{2} \sqrt{3}}
R$ for an equilateral triangle of sidelength $R$.  Figure
\ref{fig:mu0-2d} presents the small-$p$ asymptotic behavior of
$\mu_0^{(p)}$ for these four domains.  Astonishingly, the asymptotic
relation (\ref{eq:mu0-2d}) turns out to be remarkably accurate, even
though the correction term is formally of the order of
$O(1/\ln^2(\sqrt{p}))$ (e.g., $1/\ln(\sqrt{10^{-4}}) \approx -0.22$,
i.e., even at very small $p$, the correction term could still be
comparable to the leading one).  Moreover, the asymptotic relation
(\ref{eq:mu0-2d}), which was derived in \cite{Christiansen23} for
exterior domains with smooth boundary, works equally well for two
considered polygonal domains, namely the square and the equilateral
triangle.  These results suggest that the asymptotic relation
(\ref{eq:mu0-2d}) can be more general (i.e., not limited to smooth
shapes), while the correction term may be of higher order.  An
extension of Eq. (\ref{eq:mu0-2d}) to more general domains with
polygonal or Lipschitz boundary presents an interesting perspective.

\begin{figure}[t!]
\includegraphics[width=\linewidth]{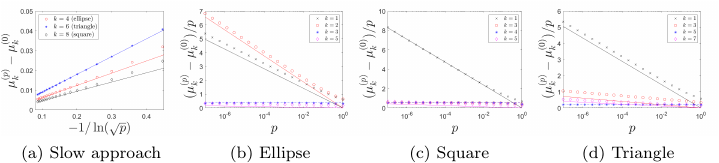}
\caption{
{\bf (a)} The small-$p$ asymptotic behavior of $\mu_k^{(p)} -
\mu_k^{(0)}$ with $k = 4$ (circles) for the exterior of an ellipse with
semiaxes $a=1$ and $b=0.5$; $k = 8$ (asterisks) for the exterior of a
square of sidelength $R=2$; and $k = 6$ (diamonds) for the exterior of
an equilateral triangle of sidelength $R=2$.  Solid lines indicate the
asymptotic relation (\ref{eq:muk_2D_conjectural}), with $a_k$ being
estimated as $0.062$ (ellipse), $0.090$ (triangle) and $0.047$
(square).
{\bf (b,c,d)} The small-$p$ asymptotic behavior of $(\mu_k^{(p)} -
\mu_k^{(0)}) / p$ for the exterior of {\bf (b)} an ellipse with
semiaxes $a=1$ and $b=0.5$, {\bf (c)} a square of sidelength $R=2$,
and {\bf (d)} an equilateral triangle of sidelength $R=2$.  Symbols
indicate the numerical results for different $k$, solid lines
present the asymptotic relation (\ref{eq:muk_2d_plogp}) when $d_k \ne
0$, and dashed lines show the asymptotic relation
(\ref{eq:muk_small_p_bounded}) when $d_k = 0$. We have
$d_1 \approx 0.62$, $d_2 \approx 0.82$, $d_3 = 0$ ($b_3 \approx
0.36$), $d_5 \approx 0.02$ (for ellipse);
$d_1 \approx 1.03$, $d_3 = d_4 = 0$ ($b_3 \approx 0.55$, $b_4 \approx
0.56$), $d_5 \approx 0.05$ (for square);
and $d_1 \approx 0.63$, $d_3 \approx 0.09$, $d_5 = 0$ ($b_5 \approx
0.21$), $d_6 \approx 0.05$ (for triangle).
We used the maximal meshsize $\hmax = 0.02$, the truncation order
$\nmax = 30$, and the outer boundary of radius $L = 2$ (see Appendix
\ref{sec:numerics} for details).}
\label{fig:mus-2d}
\end{figure}

According to our results from Sec. \ref{sec:2D_asympt}, the other
eigenvalues $\mu_k^{(p)}$ can approach their limits $\mu_k^{(0)}$ as
$O(1/\ln(\sqrt{p}))$, $O(p \ln(\sqrt{p}))$, or $O(p)$.  We observed
all three types of these asymptotic relations for the considered
examples of an ellipse, a square, and an equilateral triangle.  Figure
\ref{fig:mus-2d}(a) presents the difference $\mu_k^{(p)} -
\mu_k^{(0)}$ and the asymptotic relation (\ref{eq:muk_2D_conjectural})
with $k = 4$ (ellipse), $k = 8$ (square), and $k = 6$ (equilateral
triangle).  The coefficients $a_k$ from Eq. (\ref{eq:ck_def}) were
obtained by a numerical integration of $V_k^{(0)}$ over the circle of
radius $L$.  In turn, for other considered indices, the coefficients
$a_k$ were very close to zero, suggesting a faster approach of
$\mu_k^{(p)}$ to their limits $\mu_k^{(0)}$.
In order to check this behavior, we present on panels (b,c,d) of
Fig. \ref{fig:mus-2d} the ratio $(\mu_k^{(p)} - \mu_k^{(0)}) / p$ for
three earlier considered domains and several $k$.  For the ellipse, we
observe that $\mu_1^{(p)}$, $\mu_2^{(p)}$, and $\mu_5^{(p)}$
exhibit non-analytical small-$p$ behavior $O(p\ln(\sqrt{p}))$
according to the asymptotic relation (\ref{eq:muk_2d_plogp}), whereas
the eigenvalue $\mu_3^{(p)}$ approaches to its limit linearly with
$p$, via Eq. (\ref{eq:muk_small_p_bounded}), as for bounded domains.
After estimating the coefficients $d_k$ in front of the term
$p\ln(\sqrt{p})$ via a numerical integration in Eq. (\ref{eq:dk_def}),
we plotted the asymptotic relation (\ref{eq:muk_2d_plogp}).  These
straight lines exhibit the correct slope (given by $d_k$).  In turn,
they miss the offsets, which are determined by the next-order terms
$O(p)$ that we did not derive in Sec. \ref{sec:2D_asympt}.
A similar behavior is observed for the square (since $\mu_1^{(p)} =
\mu_2^{(p)}$ is twice degenerate, the case $k = 2$ is not shown).
Note that the eigenvalues $\mu_3^{(p)}$ and $\mu_4^{(p)}$ are distinct
but the associated coefficients $b_k$, determined by the squared
$L^2(\Omega)$-norms of their Steklov eigenfunctions $V_k^{(0)}$, turn
out to be identical.  The curve for $k = 5$ exhibits a small linear
increase due to an non-analytical leading term $O(p\ln(\sqrt{p}))$
with a small numerical prefactor $d_5 \approx 0.05$.
For the exterior of the equilateral triangle, the eigenvalues with $k
= 1, 3, 7$ exhibit an non-analytical behavior in the leading term,
whereas the eigenvalue with $k = 5$ behaves linearly with $p$.  Note
that we do not show the results for $k = 2$ and $k = 4$ because
$\mu_2^{(p)} = \mu_1^{(p)}$ and $\mu_4^{(p)} = \mu_3^{(p)}$ due to the
symmetries of the equilateral triangle.

\subsubsection*{Eigenfunctions}

\begin{figure}[t!]
\includegraphics[width=\linewidth]{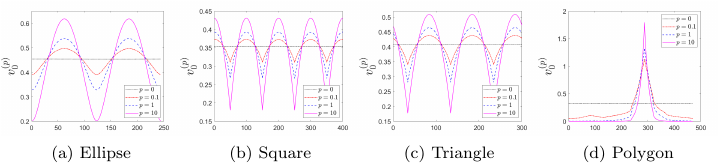}
\caption{
The principal eigenfunction $v_0^{(p)}$ with $p= 0, 0.1, 1, 10$ for
the exterior of {\bf (a)} an ellipse with semiaxes $a=1$ and $b=0.5$;
{\bf (b)} a square $(-1,1)^2$ of sidelength $R=2$; {\bf (c)} an
equilateral triangle of sidelength $R=2$; and {\bf (d)} the polygonal
shape shown in Fig. \ref{fig:V0-2d}(bottom row).  The horizontal axis
corresponds to a curvilinear coordinate along the boundary $\pa$ (the
index of equidistant boundary points), which follows the polar angle
from $-\pi$ to $\pi$.  For panels (a,b,c), the starting index
corresponds to the boundary point with the polar angle $-\pi$, i.e.,
the point $(-1,0)$ for the ellipse and the square, the point $(-0.67,
0)$ for the equilateral triangle, and the point $(1, -1)$ for the
polygonal shape.  Note that four and three spikes on panels (b) and
(c) correspond to the vertices of the square and of the triangle,
respectively.
We used the maximal meshsize $\hmax = 0.02$, the truncation
order $\nmax = 30$, and the outer boundary of radius $L = 2$ (see
Appendix \ref{sec:numerics} for details).  }
\label{fig:v0-2d}
\end{figure}

\begin{figure}[t!]
\centering
\includegraphics[width=0.8\linewidth]{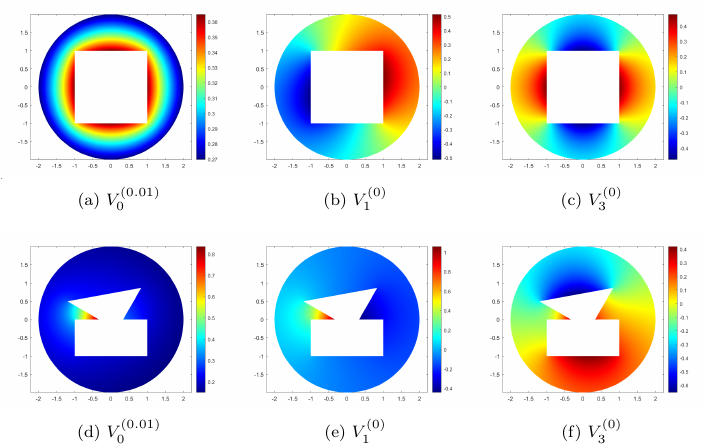}
\caption{
Several Steklov eigenfunctions $V_k^{(p)}$ for the exterior of a
square of sidelength $R = 2$ (top) and of a polygonal domain (bottom).
We used the maximal meshsize $\hmax = 0.02$, the truncation order
$\nmax = 30$, and the outer boundary of radius $L = 2$ (see Appendix
\ref{sec:numerics} for details).}
\label{fig:V0-2d}
\end{figure}

We inspect the behavior of the principal eigenfunction $v_0^{(p)}$ of
the Dirichlet-to-Neumann operator $\M_p$ for different shapes and
values of $p$.  Figure \ref{fig:v0-2d} illustrates the validity of
Eq. (\ref{eq:v0-2d}) for a wide range of shapes such as the exterior
of an ellipse, a square, an equilateral triangle and a polygonal shape
shown in Fig. \ref{fig:V0-2d}(bottom row).  In contrast to the
exterior of a disk, $v_0^{(p)}$ depends on $p$ for these shapes.
Moreover, as $p$ increases, the eigenfunction exhibits variations of
larger and larger amplitudes.  For the last domain, $v_0^{(p)}$
becomes localized near the vertex with the smallest angle $\pi/6$.
Note that the shown principal eigenfunctions $v_0^{(p)}$ are positive
for all the considered cases.  While this property is known for the
interior Steklov problem \cite{Arendt07,Arendt12}, its extension to
exterior domains remains an open question.

Since $v_0^{(0)}$ is constant, the related Steklov eigenfunction
$V_0^{(0)}$ is also constant over the whole domain $\Omega$ (not
shown).  In turn, as soon as $p > 0$, the principal Steklov
eigenfunction $V_0^{(p)}(\x)$ exhibits an exponential decay at
infinity (as $|\x|\to\infty$).  Panels (a) and (d) of
Fig. \ref{fig:V0-2d} illustrate this behavior for the exterior of a
square and of a polygonal domain.  Two other Steklov eigenfunctions
$V_1^{(0)}$ and $V_3^{(0)}$ are also illustrated in
Fig. \ref{fig:V0-2d}.  Both $V_0^{(0.01)}$ and $V_1^{(0)}$ are
localized in the angle $\pi/6$ of the polygonal domain, in analogy to
the interior problem (see \cite{Chaigneau24} and references therein).

\begin{figure}[t!]
\centering
\includegraphics[width=0.9\linewidth]{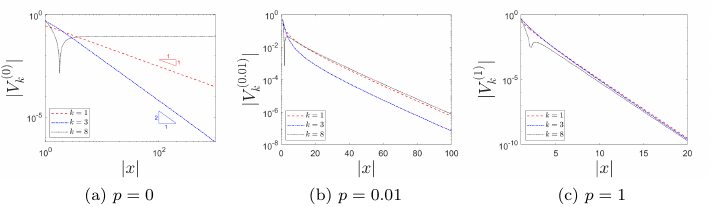}
\caption{
Decay of three Steklov eigenfunctions $|V_k^{(p)}|$ along a vertical
line for the exterior of a square of sidelength $R=2$, for three
values of $p$.  The associated eigenvalues are: $\mu_1^{(0)} \approx
0.76$, $\mu_3^{(0)} \approx 1.30$, $\mu_8^{(0)} \approx 3.37$;
$\mu_1^{(0.01)} \approx 0.79$, $\mu_3^{(0.01)} \approx 1.31$,
$\mu_8^{(0.01)} \approx 3.40$; and $\mu_1^{(1)} \approx 1.43$,
$\mu_3^{(1)} \approx 1.70$, $\mu_8^{(1)} \approx 3.62$.  A ``spike''
for the dashed curve corresponds to the sign change of the
eigenfunction $V_8^{(0)}$.  We used the maximal meshsize $\hmax =
0.01$, the truncation order $\nmax = 30$, and the outer boundary of
radius $L = 2$ (see Appendix \ref{sec:numerics} for details).  }
\label{fig:Vk-2dext}
\end{figure}

Figure \ref{fig:Vk-2dext} shows the decay of Steklov eigenfunctions
$V_k^{(0)}$ for the exterior of a square.  For this purpose, we plot
$|V_k^{(0)}|$ along the negative part of the vertical axis.  One sees
that the eigenfunction with $k = 1$ decays as $1/|\x|$ so that their
$L^2(\Omega)$-norm is infinite, yielding an non-analytic asymptotic
behavior of the associated eigenvalues $\mu_k^{(p)}$, as shown in
Fig. \ref{fig:mus-2d}.  In turn, the eigenfunction $V_3^{(0)}$ decays
as $1/|\x|^2$ so that its $L^2(\Omega)$-norm is finite, and the
associated eigenvalue $\mu_3^{(p)}$ approaches linearly to its limit
$\mu_3^{(0)}$.  Finally, the eigenfunction $V_8^{(0)}$ does not decay
but approaches a constant, as in Eq. (\ref{eq:Vk_inf}), that agrees
with the asymptotic behavior (\ref{eq:muk_2D_conjectural}) of the
eigenvalue $\mu_8^{(p)}$, as observed in Fig. \ref{fig:mus-2d}(a).

\subsection{Three dimensions}
\label{sec:3D}

A numerical analysis of the exterior Steklov problem in three
dimensions can be carried out by the same finite-element method with
the transparent boundary condition.  However, a practical
implementation of three-dimensional meshing and the consequent
numerical diagonalization of large matrices can be rather
time-consuming.  For our illustrative purposes, it is sufficient to
consider axisymmetric domains, for which the original
three-dimensional problem can be treated as effectively
two-dimensional (see Appendix \ref{sec:numerics} for details).
Moreover, we focus here exclusively on axisymmetric eigenfunctions,
i.e., throughout this subsection, $V_k^{(p)}$ denotes the $(k+1)$-th
axisymmetric Steklov eigenfunction, while $\mu_k^{(p)}$ is the
associated eigenvalue.  Other eigenvalues and eigenfunctions are not
discussed (see \cite{Grebenkov24} for a detailed study of various
Steklov eigenfunctions with $p = 0$ for prolate and oblate spheroids).

\subsubsection*{Eigenvalues}

\begin{figure}[t!]
\includegraphics[width=\linewidth]{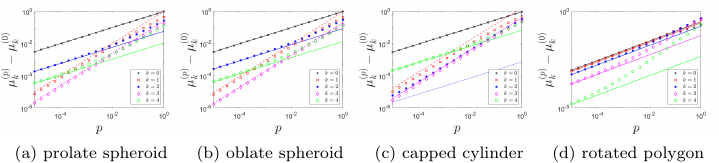}
\caption{
Small-$p$ asymptotic behavior of $\mu_k^{(p)} - \mu_k^{(0)}$ for the
exterior of {\bf (a)} a prolate spheroid with semiaxes $a=0.5$, $b=1$;
{\bf (b)} an oblate spheroid with semiaxes $a=0.5$, $b=1$; {\bf (c)} a
capped cylinder of radius $1$ and height $2$; and {\bf (d)} a domain
obtained by rotating the polygon shown in
Fig. \ref{fig:Vk-3d}(bottom).  Symbols show the numerical results (as
indicated in the legend), whereas lines of the same color indicate the
associated asymptotic behavior: solid lines for
Eq. (\ref{eq:muk_ak_3d}) and dashed lines for
Eq. (\ref{eq:muk_small_p_bounded}).  A numerical computation of the
coefficient $b_k$ given by Eq. (\ref{eq:bk}) is described in Appendix
\ref{sec:bk}.  We used the maximal meshsize $\hmax = 0.02$, the truncation
order $\nmax = 30$, and the outer boundary of radius $L = 2$ (see
Appendix \ref{sec:numerics} for details).}
\label{fig:mu0-3d}
\end{figure}

Figure \ref{fig:mu0-3d} illustrates the small-$p$ asymptotic behavior
of first eigenvalues $\mu_k^{(p)}$ for the exterior of prolate/oblate
spheroids, of a capped cylinder, and of a domain obtained rotating a
given polygon around $z$ axis.  For the case of prolate/oblate
spheroids, the axisymmetric Steklov eigenfunctions $V_k^{(0)}$ were
shown to be symmetric (resp., antisymmetric) with respect to the
horizontal plane when $k$ is even (resp., odd) \cite{Grebenkov24}.  As
a consequence, the integral of $v_k^{(0)}$ over the boundary $\pa$ is
zero for odd indices $k$, implying $a_k = 0$ and the linear asymptotic
relation (\ref{eq:muk_small_p_bounded}).  In turn, these integrals are
nonzero for even indices $k$, yielding the asymptotic relation
(\ref{eq:muk_ak_3d}), in agreement with our numerical results shown on
panels (a) and (b).  Similarly, the symmetries of the capped cylinder
imply that the integrals of $v_k^{(0)}$ over the boundary $\pa$ vanish
for the eigenfunctions with $k=1$ and $k=3$.  In turn, we obtain the
asymptotic behavior (\ref{eq:muk_ak_3d}) with $a_0 \approx 0.91$, $a_2
\approx 7.2 \cdot 10^{-4}$ and $a_4 \approx 0.07$ for the associated
eigenvalues.  Since the numerical value of $a_2$ is quite small, one
needs to reach much smaller $p$ to ensure that $\sqrt{p}$ is indeed
the leading term.  In other words, the smallness of $a_2$ explains why
the asymptotic relation (\ref{eq:muk_ak_3d}) is not yet achieved at
the considered values of $p$.
Finally, for a general domain (without symmetries), there is no reason
to get $a_k = 0$ for axisymmetric eigenfunctions, as illustrated on
panel (d).

\subsubsection*{Eigenfunctions}

\begin{figure}[t!]
\centering
\includegraphics[width=0.8\linewidth]{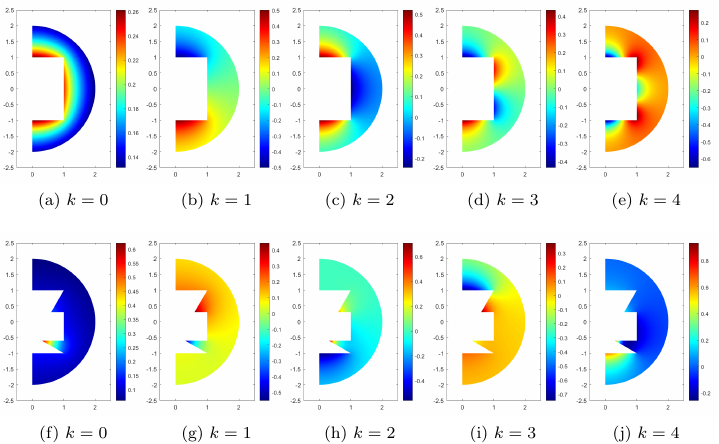}
\caption{
First axisymmetric Steklov eigenfunctions $V_k^{(0)}$ for the exterior
of {\bf (top)} a capped cylinder of radius $1$ and height $2$; {\bf
(bottom)} an axisymmetric domain obtained by rotating the shown
polygon along the vertical axis.  The projection onto $xz$ plane,
corresponding to the angular coordinate $\phi = 0$, is shown.  We used
the maximal meshsize $\hmax = 0.02$, the truncation order $\nmax
= 30$, and the outer boundary of radius $L = 2$ (see Appendix
\ref{sec:numerics} for details).}
\label{fig:Vk-3d}
\end{figure}

Top panels of Fig. \ref{fig:Vk-3d} present first five axisymmetric
Steklov eigenfunctions $V_k^{(0)}$ for the exterior of a capped
cylinder of radius $1$ and height $2$.  The shown projection on the
$xz$ plane fully captures their behavior in the whole domain due to
their axisymmetric symmetry.  As for interior domains
\cite{Arendt07,Arendt12}, the principal eigenfunction $V_0^{(0)}$ is
positive, while the other eigenfunctions change their sign.  One sees
that the eigenfunctions $V_1^{(0)}$ and $V_3^{(0)}$ are antisymmetric
with respect to the horizontal plane $xy$ (or, equivalently, to the
horizontal axis for the shown projection) so that the integral of
$v_k^{(0)}$ over the boundary $\pa$ is strictly zero, yielding $a_1 =
a_3 = 0$, in agreement with the small-$p$ behavior shown in
Fig. \ref{fig:mu0-3d}(c).  In turn, the shown eigenfunctions
$V_0^{(0)}$, $V_2^{(0)}$ and $V_4^{(0)}$ are symmetric with respect to
the horizontal plane, suggesting that $a_k > 0$, in agreement with the
previously reported values of $a_k$.  

Bottom panels of Fig. \ref{fig:Vk-3d} present first five axisymmetric
Steklov eigenfunctions $V_k^{(0)}$ for the exterior of an axisymmetric
domain obtained by rotating the shown polygon along the vertical axis.
This polygon was chosen to have two obtuse angles $2\pi - \pi/3$ and
$2\pi -\pi/6$ whose complementary angles $\pi/3$ and $\pi/6$ may
accommodate localized Steklov eigenfunctions, e.g., $V_0^{(0)}$ on
Fig. \ref{fig:Vk-3d}(f).  Expectedly, a somewhat arbitrary shape of
this polygon eliminates most symmetries (except for the axial symmetry
imposed by construction) in the shown eigenfunctions.  In particular,
the integral of $v_k^{(0)}$ over $\pa$ is not zero for all indices $k
\in\{ 0,1,2,3,4\}$, yielding $a_k > 0$ and thus the asymptotic
behavior (\ref{eq:muk_ak_3d}) shown in Fig. \ref{fig:mu0-3d}(d).

\begin{figure}[t!]
\includegraphics[width=0.9\linewidth]{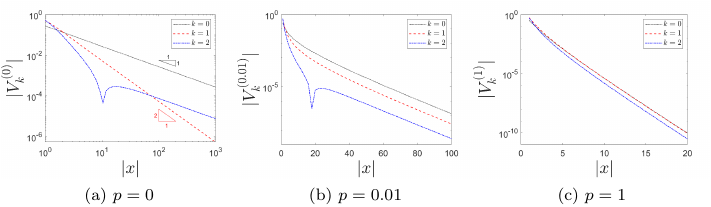}
\caption{
Decay of three Steklov eigenfunctions $|V_k^{(p)}|$ along a vertical
line for the exterior of a capped cylinder of radius $1$ and height
$2$, for three values of $p$.  Note that a spike for blue dash-dotted
line ($k = 2$) on panels (a) and (b) corresponds to a sign change of
$V_k^{(p)}$.
The associated eigenvalues are: $\mu_0^{(0)} \approx 0.78$,
$\mu_1^{(0)} \approx 1.47$, $\mu_2^{(0)} \approx 1.96$;
$\mu_0^{(0.01)} \approx 0.87$, $\mu_1^{(0.01)} \approx 1.48$,
$\mu_2^{(0.01)} \approx 1.96$;
$\mu_0^{(1)} \approx 1.70$, $\mu_1^{(1)} \approx 1.95$, $\mu_2^{(1)}
\approx 2.27$. 
We used the maximal meshsize $\hmax = 0.02$, the truncation order
$\nmax = 30$, and the outer boundary of radius $L = 2$ (see Appendix
\ref{sec:numerics} for details).  }
\label{fig:Vk-pro-3d}
\end{figure}

After examination of spatial variations of the Steklov functions near
the boundary, we turn to their asymptotic behavior at large $|\x|$.
Figure \ref{fig:Vk-pro-3d} illustrates the expected power-law decay of
Steklov eigenfunctions for $p=0$ and the exponential decay
(\ref{eq:Vk_bound_exp}) of Steklov eigenfunctions for $p>0$ for the
exterior of a capped cylinder.  Since $a_k > 0$ for $k = 0$ and $k =
2$, the associated eigenfunctions $|V_k^{(0)}|$ decreases as $1/|\x|$,
according to Eq. (\ref{eq:Vk0_asympt}) with $\gamma_k = d-2 = 1$.  In
turn, as $a_2 = 0$, the eigenfunction $V_2^{(0)}$ shows a faster
power-law decay (here, with the exponent $-2$).  When $p > 0$, all
Steklov eigenfunctions exhibit an exponential decay, with an upper
bound decreasing as $e^{-\sqrt{p} |\x|}$ (see Appendix
\ref{sec:upper}).  This behavior is confirmed on panels (b) and (c) of
Fig. \ref{fig:Vk-pro-3d}.

\section{Statistics of first-passage times}
\label{sec:application}

In this section, we describe an application of the small-$p$
asymptotic results to diffusion-controlled reactions and related
first-passage time statistics.  Indeed, as the modified Helmholtz
equation $(p-\Delta) \tilde{u} = 0$ is related to the diffusion
equation $\partial_t u = \Delta u$ via the Laplace transform, the
small-$p$ behavior of $\tilde{u}(\x,p)$ corresponds to the long-time
behavior of $u(\x,t)$.  In particular, the encounter-based approach
allowed one to express most characteristics of diffusion-reaction
processes in the Laplace domain in terms of the Steklov eigenfunctions
and eigenvalues
\cite{Grebenkov19,Grebenkov19c,Grebenkov20,Grebenkov20b,Grebenkov20c,Grebenkov22b,Grebenkov22d,Grebenkov22a}.

To illustrate this point, we focus on the distribution of the
first-crossing time $\T_\ell = \inf\{ t > 0 ~:~ \ell_t > \ell\}$,
i.e., the first time instance when the boundary local time $\ell_t$
crosses a fixed threshold $\ell$.  The stochastic process $\ell_t$
naturally appears in the Skorokhod (or Langevin) stochastic equation
of reflected Brownian motion in the presence of a reflecting boundary
\cite{Skorokhod61,Skorokhod62,Ito,Freidlin,Saisho87} and represents
the (rescaled) number of encounters between the diffusing particle and
the boundary $\pa$ up to time $t$, and thus characterizes surface
reactions \cite{Grebenkov19c}.  In turn, the random variable $\T_\ell$
plays the central role in defining the first-reaction times
\cite{Grebenkov20,Grebenkov20b}.  The moment-generating function of the
first-crossing time was expressed in \cite{Grebenkov20} as
\begin{equation}   \label{eq:tildeU}
\tilde{U}(\ell,p|\x_0) = \langle e^{-p \T_\ell} \rangle = \int\limits_0^\infty dt \, e^{-pt} \, U(\ell,t|\x_0) 
 = \sum\limits_{k=0}^\infty V_k^{(p)}(\x_0) e^{-\mu_k^{(p)}\ell} \int\limits_{\pa} v_k^{(p)} ,
\end{equation}
where $\x_0$ is the starting point, and $U(\ell,t|\x_0)$ is the
probability density function (PDF) of $\T_\ell$.
For simplicity, we assume that the particle starts on the boundary
$\pa$ (if $\x_0 \notin \pa$, there is an additional stage of the first
arrival onto $\pa$; as this first-passage process was thoroughly
investigated in the past \cite{Redner,Metzler,Lindenberg,Grebenkov},
we skip it here and set directly $\x_0\in\pa$).

Let us deduce the small-$p$ asymptotic behavior of
Eq. (\ref{eq:tildeU}) in three dimensions.  In the leading order, one
can replace $v_k^{(p)}$ by its limit $v_k^{(0)}$.  As the resulting
expression includes the integrals of $v_k^{(0)}$ over $\pa$, only the
eigenmodes with nonzero integrals do contribute in the leading order.
One can therefore use the asymptotic relation (\ref{eq:muk_ak_3d}),
with $a_k$ given by Eq. (\ref{eq:ak}) so that
\begin{equation} 
\tilde{U}(\ell,p|\x_0) \approx \sum\limits_{k=0}^\infty v_k^{(0)}(\x_0) e^{-(\mu_k^{(0)} + a_k \sqrt{p})\ell} \int\limits_{\pa} v_k^{(0)} 
\qquad (p\to 0),
\end{equation}
from which the inverse Laplace transform yields
\begin{equation}  \label{eq:Ut_asympt}
U(\ell,t|\x_0) \approx \sum\limits_{k=0}^\infty \frac{a_k \ell}{\sqrt{4\pi Dt^3}} e^{-\mu_k^{(0)}\ell-a_k^2 \ell^2/(4Dt)}
v_k^{(0)}(\x_0) \int\limits_{\pa} v_k^{(0)}  \qquad (t\to \infty),
\end{equation}
where we included the diffusion coefficient $D$ (which was set to $1$
in the remaining text).  For instance, for a sphere of radius $L$, one
has $\mu_0^{(0)} = 1/L$, $a_0 = 1$ and $v_0^{(0)} = 1/\sqrt{4\pi
L^2}$, whereas the other eigenfunctions $v_k^{(0)}$ are orthogonal to
$v_0^{(0)}$, yielding
\begin{equation}
U(\ell,t|\x_0) = \frac{\ell \, e^{-\ell/L - \ell^2/(4Dt)}}{\sqrt{4\pi Dt^3}}   \qquad (\x_0\in\pa).
\end{equation}
Note that, in the case of the sphere, this expression turns out to be
exact, i.e., it is valid for any $t$ (see \cite{Grebenkov21a}).

The explicit form of the long-time asymptotic relation
(\ref{eq:Ut_asympt}) highlights the relative roles of the eigenvalues
$\mu_k^{(0)}$ and the coefficients $a_k$.  In the common setting of a
partially reactive boundary with a constant reactivity parameter $q$,
the fixed threshold $\ell$ is replaced by a random threshold
$\hat{\ell}$ with the exponential distribution, $\P\{ \hat{\ell} >
\ell\} = e^{-q\ell}$ (see \cite{Grebenkov19,Grebenkov20}).  As a
consequence, the probability density function $H_q(t|\x_0)$ of the
first-reaction time is obtained by averaging $U(\hat{\ell},t|\x_0)$
over all random realizations of $\hat{\ell}$:
\begin{equation}
H_q(t|\x_0) = \langle U(\hat{\ell},t|\x_0) \rangle = \int\limits_0^\infty d\ell \,
\underbrace{qe^{-q\ell}}_{\textrm{PDF of}~\hat{\ell}} \, U(\ell,t|\x_0).
\end{equation}
Using the asymptotic relation (\ref{eq:Ut_asympt}), we get the
long-time behavior (for $\x_0 \in \pa$):
\begin{equation}  \label{eq:Hq_asympt}
H_q(t|\x_0) \approx qD \sum\limits_{k=0}^\infty \biggl\{\frac{1}{\sqrt{\pi Dt}} - \frac{\mu_k^{(0)}+q}{a_k} \,
\erfcx\biggl(\sqrt{Dt} \, \frac{\mu_k^{(0)}+q}{a_k}\biggr)\biggr\} \frac{v_k^{(0)}(\x_0)}{a_k}  \int\limits_{\pa} v_k^{(0)}  ,
\end{equation}
where $\erfcx(z) = e^{z^2} \erfc(z)$ is the scaled complementary error
function.  For the sphere of radius $L$, only the principal eigenmode
contributes (with $a_0 = 1$ and $\mu_0^{(0)} = 1/L$), yielding
\begin{equation}  \label{eq:Hq_sphere}
H_q(t|\x_0) = qD \biggl\{\frac{1}{\sqrt{\pi Dt}} - (1/L+q) \, \erfcx\biggl(\sqrt{Dt} \, (1/L+q)\biggr)\biggr\}  \qquad (\x_0\in\pa) ,
\end{equation}
and this asymptotic relation turns out to be exact (see
\cite{Grebenkov18} for details).  
To our knowledge, such a detailed representation (\ref{eq:Hq_asympt})
of the long-time asymptotic behavior of the PDF $H_q(t|\x_0)$ was not
earlier reported for general exterior domains in three dimensions.

At very long times, one can use the asymptotic behavior of $\erfcx(z)$
as $z\to \infty$ to get
\begin{equation}
H_q(t|\x_0) \approx \frac{q}{\sqrt{4\pi Dt^3}} \sum\limits_{k=0}^\infty 
\frac{a_k}{(\mu_k^{(0)}+q)^2} v_k^{(0)}(\x_0) \int\limits_{\pa} v_k^{(0)}  .
\end{equation}
Expectedly, one retrieves the characteristic $t^{-3/2}$ heavy tail of
the PDF in three dimensions, whereas the coefficients $a_k$ and
eigenfunctions $v_k^{(0)}$ determine the prefactor.

Finally, the integral of $H_q(t|\x_0)$ from $t$ to infinity determines
the survival probability $S_q(t|\x_0)$ of the particle in the presence
of a partially reactive boundary $\pa$.  Substituting
Eq. (\ref{eq:Hq_asympt}) for large $t$, one gets the long-time
asymptotic behavior of this survival probability (for $\x_0\in\pa$),
\begin{align}  \nonumber
S_q(t|\x_0) & \approx \sum\limits_{k=0}^\infty \biggl\{1 - \frac{q}{\mu_k^{(0)} + q}\biggl[1 -  
\erfcx\biggl(\sqrt{Dt}\, \frac{\mu_k^{(0)}+q}{a_k}\biggr)\biggr]\biggr\} v_k^{(0)}(\x_0) \int\limits_{\pa} v_k^{(0)}  \\  \label{eq:Sq_asympt}
& \approx 1 - q \sum\limits_{k=0}^\infty \biggl[1 -  
\erfcx\biggl(\sqrt{Dt}\, \frac{\mu_k^{(0)}+q}{a_k}\biggr)\biggr] \frac{v_k^{(0)}(\x_0)}{\mu_k^{(0)} + q}  \int\limits_{\pa} v_k^{(0)}  ,
\end{align}
where we used the completeness of the eigenfunctions $v_k^{(0)}$ to
evaluate explicitly the first term (i.e., the constant $1$).  For the
sphere, this asymptotic relation contains only the eigenmode with $k =
0$ and turns out to be exact, reproducing the seminal result by
Collins and Kimball \cite{Collins49}.  In the limit $t\to\infty$, the
asymptotic relation (\ref{eq:Sq_asympt}) yields the exact form of the
limiting survival probability (for $\x_0\in\pa$):
\begin{equation}  
S_q(\infty|\x_0) = 1 - q \sum\limits_{k=0}^\infty 
\frac{v_k^{(0)}(\x_0)}{\mu_k^{(0)} + q} \int\limits_{\pa} v_k^{(0)}  .
\end{equation}
This probability is nonzero due to the transient character of Brownian
motion in three dimensions.  In particular, $1 - S_q(\infty|\x_0)$ is
the probability of escape to infinity.
Moreover, even though Eq. (\ref{eq:Sq_asympt}) was derived in the
long-time limit, its right-hand side approaches the correct limit $1$
as $t \to 0$.  One can therefore expect that this relation might be a
good approximation for the survival probability $S_q(t|\x_0)$ over the
whole range of times.  A numerical validation of this approximation
presents an interesting perspective.

Note that a similar analysis can be performed for the probability
density function of the boundary local time
\cite{Grebenkov19c,Grebenkov20}.  Moreover, one can also investigate
the long-time asymptotic behavior of $U(\ell,t|\x_0)$, $H_q(t|\x_0)$
and $S_q(t|\x_0)$ in two dimensions (see \cite{Grebenkov21a} for the
related analysis in the case of the exterior of a disk).

\section{Conclusion}
\label{sec:conclusion}

In this paper, we studied the small-$p$ asymptotic behavior of the
eigenvalues and eigenfunctions of the generalized exterior Steklov
problem.  The exponential decay of the Steklov eigenfunctions
$V_k^{(p)}$ at infinity allowed us to explore the spectral properties
for any $p > 0$ and then to consider the limit $p\to 0$.  In
particular, we derived the simple relation (\ref{eq:dmu}) between the
squared $L^2(\Omega)$-norm of the eigenfunction $V_k^{(p)}$ and the
derivative of the associated eigenvalue $\mu_k^{(p)}$ with respect to
$p$.  In the limit $p\to 0$, this relation distinguishes two
situations: (i) when the $L^2(\Omega)$-norm of the limiting
eigenfunction $V_k^{(0)}$ is finite, the associated eigenvalue
$\mu_k^{(p)}$ approaches its limit $\mu_k^{(0)}$ linearly with $p$, as
in the case of bounded domains; (ii) when the norm is infinite, an
non-analytic behavior of $\mu_k^{(p)}$ is expected near $p = 0$.  We
managed to identify the leading non-analytic term for exterior domains
in different space dimensions.  In two dimensions, the principal
eigenvalue $\mu_0^{(p)}$ vanishes logarithmically, as earlier proved
rigorously in \cite{Christiansen23}, while the associated
eigenfunction becomes constant.  This behavior was extended via
Eq. (\ref{eq:muk_2D_conjectural}) to other eigenvalues, whose
eigenfunctions $V_k^{(0)}$ do not vanish at infinity (and thus the
integrals $c_k$ defined by Eq. (\ref{eq:ck_def}) were not zero).
Numerical examples of such behavior were shown in
Fig. \ref{fig:mus-2d}(a).  In turn, when $c_k = 0$ but the
$L^2(\Omega)$-norm of $V_k^{(0)}$ is still infinite, the difference
$\mu_k^{(p)} - \mu_k^{(0)}$ exhibits the non-analytic behavior
$p\ln(\sqrt{p})$, which was observed for various exterior domains.
Finally, if the $L^2(\Omega)$-norm of $V_k^{(0)}$ is finite, one
retrieves the conventional linear relation
(\ref{eq:muk_small_p_bounded}), as for bounded domains.
In three dimensions, the finiteness of the $L^2(\Omega)$-norm of
$V_k^{(0)}$ is directly related to vanishing of the integral of the
eigenfunction $v_k^{(0)}$ over the boundary $\pa$: if this integral is
zero, the norm is finite, Eq. (\ref{eq:dmu}) holds, and $\mu_k^{(p)} -
\mu_k^{(0)}$ is proportional to $p$ in the leading order.  However, in
the generic situation when the integral is not zero, the leading term
becomes $\sqrt{p}$ in Eq. (\ref{eq:muk_ak_3d}), with the coefficient
$a_k$ given by Eq. (\ref{eq:ak}).  In the same vein, in four
dimensions, if the integral of $v_k^{(0)}$ over $\pa$ is not zero, the
non-analytic leading term $p\ln(\sqrt{p})$ emerges.  In turn, all
eigenfunctions $V_k^{(0)}$ have a finite $L^2(\Omega)$-norm in higher
dimensions, implying the linearity of $\mu_k^{(p)} - \mu_k^{(0)}$ with
$p$ in the leading order.
These theoretical results were illustrated by several examples of
exterior domains in two and three dimensions.  For this purpose, we
modified the earlier developed FEM by introducing the transparent
boundary condition to deal with an equivalent Steklov problem in a
bounded domain.

From the mathematical point of view, these asymptotic results clarify
some properties of the exterior Steklov problem in the limit $p = 0$.
For instance, this problem appears to be well posed in the
conventional space $H^1(\Omega)$ in high dimensions $d \geq 5$.  In
turn, in dimensions $d = 2,3,4$, the $L^2(\Omega)$-norm of at least
one limiting eigenfunction $V_k^{(0)}$ is expected to be infinite,
thus requiring either a larger functional space
\cite{Auchmuty14,Auchmuty18,Xiong23}, or a sequence of bounded
domains approaching the unbounded one \cite{Arendt15}.  A mathematical
proof of the existence of the limit $p = 0$ and the inter-connections
between various formulations of the exterior Steklov problem will be
discussed in a separate work \cite{Bundrock25}.

From the physical point of view, the small-$p$ asymptotic behavior of
the eigenvalues and eigenfunctions of the generalized exterior Steklov
problem has immediate consequences for diffusion-controlled reactions
and related first-passage time statistics.  We derived the long-time
asymptotic behavior of the PDF $U(\ell,t|\x_0)$ of the first-crossing
time in three dimensions.  Here the integrals of eigenfunctions
$v_k^{(0)}$ over the boundary $\pa$ determine which Steklov eigenmodes
do contribute to this asymptotic behavior, while the related
coefficients $a_k$ control the exponential decay of these
contributions.  This PDF determines the statistics of various
first-reaction times.  In particular, we deduced the long-time
asymptotic behavior of the survival probability and the related PDF of
the first-reaction time in the common setting of a partially reactive
boundary with constant reactivity.  These asymptotic results emphasize
the role of the coefficients $a_k$ and open a way to quantify the
impact of the geometric shape of the boundary onto
diffusion-controlled reactions in exterior domains.

\section*{Acknowledgments}

The authors thank professors I. Polterovich and M. Levitin for
fruitful discussions.  D.S.G. acknowledges the Simons Foundation for
supporting his sabbatical sojourn in 2024 at the CRM, University of
Montr\'eal, Canada.  D.S.G. also acknowledges the Alexander von
Humboldt Foundation for support within a Bessel Prize award.

\section*{Data Availability Statement}

The data that support the findings of this study are available from
the corresponding author upon reasonable request.

\appendix
\section{Upper bounds on Steklov eigenfunctions}
\label{sec:upper}

In this Appendix, we inspect the decay of Steklov eigenfunctions at
infinity.  While this behavior is known, we employ some probabilistic
arguments to get upper bounds.

As previously, we consider the exterior of a compact set $\Omega_0$,
$\Omega = \R^d \backslash \Omega_0$, with a smooth boundary $\pa$.
Each Steklov eigenfunction can be written as an extension of the
corresponding eigenfunction $v_k^{(p)}$, defined on $\pa$:
\begin{equation}
V_k^{(p)}(\x_0) = \int\limits_{\pa} d\x \, v_k^{(p)}(\x) \, \tilde{j}_\infty(\x,p|\x_0),
\end{equation} 
where $\tilde{j}_\infty(\x,p|\x_0) = \int\nolimits_0^\infty dt \,
e^{-pt} \, j_\infty(\x,t|\x_0)$ is the Laplace transform of the
probability flux density $j_\infty(\x,t|\x_0) = - \partial_n
G_\infty(\x,t|\x_0)$ (hereafter tilde denotes the Laplace transform).
Here $G_\infty(\x,t|\x_0)$ is the heat kernel (or propagator),
satisfying the diffusion equation, $\partial_t G_\infty(\x,t|\x_0) =
\Delta G_\infty(\x,t|\x_0)$, with Dirichlet condition condition,
$G_\infty(\x,t|\x_0)|_{\pa} = 0$, and the initial condition
$G_\infty(\x,0|\x_0) =\delta(\x-\x_0)$ with a Dirac distribution
$\delta(\x-\x_0)$.  Note that the diffusion coefficient is set to $1$
for brevity.

As a consequence, one gets
\begin{equation}
|V_k^{(p)}(\x_0)| \leq \| v_k^{(p)}\|_{L^\infty(\pa)} \, \tilde{H}_\infty(p|\x_0) ,
\end{equation} 
where
\begin{equation}
\tilde{H}_\infty(p|\x_0) = \int\limits_{\pa} d\x \, \tilde{j}_\infty(\x,p|\x_0)
\end{equation}
is the Laplace transform of the probability density $H_\infty(t|\x_0)$
of the first-passage time $\tau$ to $\pa$, whereas $\|
v_k^{(p)}\|_{L^\infty(\pa)}$ is the maximum of $|v_k^{(p)}|$ over
$\pa$.  Note that the probability density $H_\infty(t|\x_0)$ is not
normalized to $1$ for $d \geq 3$ due to the transient character of
Brownian motion.

Let $\Omega_0 \subset \Omega^\star_0$ so that $\Omega^\star = \R^d\backslash
\Omega^\star_0 \subset \Omega$.  For a starting point $\x_0 \in \Omega^\star$,
we introduce the first-passage time $\tau^\star$ to the boundary $\pa^\star$.
Since the trajectories of Brownian motion are continuous, each
trajectory arriving to $\pa$ must cross $\pa^\star$.  As a consequence, one
can formally write $\tau^\star \leq \tau$, which means that
\begin{equation}
S^\star_\infty(t|\x_0) = \P_{\x_0}\{ t < \tau^\star\} \leq \P_{\x_0}\{ t < \tau\} = S_\infty(t|\x_0),
\end{equation}
where $S_\infty(t|\x_0)$ and $S^\star_\infty(t|\x_0)$ are the survival
probabilities in $\Omega$ and $\Omega^\star$, respectively.  Multiplying
this inequality by $e^{-pt}$ and integrating over $t$ from $0$ to
infinity, one gets for the Laplace transforms:
\begin{equation}
\tilde{S}^\star_\infty(p|\x_0) \leq \tilde{S}_\infty(p|\x_0) \qquad (p \geq 0).
\end{equation}
In turn, as $H_\infty(t|\x_0) = -\partial_t S_\infty(t|\x_0)$ and
$H^\star_\infty(t|\x_0) = -\partial_t S^\star_\infty(t|\x_0)$, one has
\begin{equation}
\tilde{H}_\infty(p|\x_0) = 1 - p \tilde{S}_\infty(p|\x_0) \leq 1 - p \tilde{S}^\star_\infty(p|\x_0) = \tilde{H}^\star_\infty(p|\x_0).
\end{equation}
We conclude that $\tilde{H}^\star_\infty(p|\x_0)$ plays the role of an
upper bound for $\tilde{H}_\infty(p|\x_0)$ and thus for
$|V_k^{(p)}(\x_0)|$:
\begin{equation}  \label{eq:Vk_estimate}
|V_k^{(p)}(\x_0)| \leq \| v_k^{(p)}\|_{L^\infty(\pa)} \, \tilde{H}^\star_\infty(p|\x_0) .
\end{equation} 

Choosing $\Omega^\star_0 = \{ \x\in \R^d ~:~ |\x| < L\}$ to be a ball
of radius $L$ that englobes $\Omega_0$, one has
\begin{equation}
\tilde{H}^\star_\infty(p|\x_0) = (|\x_0|/L)^{1-\frac{d}{2}} \frac{K_{\frac{d}{2}-1}(|\x_0|\sqrt{p})}{K_{\frac{d}{2}-1}(L\sqrt{p})} \,,
\end{equation}
where $K_\nu(z)$ is the modified Bessel function of the second kind.
For $p > 0$ and large $|\x_0|$, one uses the asymptotic behavior of
$K_\nu(z)$ to get
\begin{equation}\label{eq:Jprime_p}
\tilde{H}^\star_\infty(p|\x_0) \approx (|\x_0|/L)^{\frac{1-d}{2}}  \sqrt{\frac{\pi}{2}}  
\frac{(L^2 p)^{-1/4}}{K_{\frac{d}{2}-1}(L\sqrt{p})}\, e^{-|\x_0|\sqrt{p}} \,.
\end{equation}
As a consequence, each Steklov eigenfunction $V_k^{(p)}(\x_0)$ decays
exponentially fast at infinity for any $p > 0$:
\begin{equation}  \label{eq:Vk_bound_exp}
|V_k^{(p)}(\x_0)| \leq C_{k,p} \, e^{-|\x_0|\sqrt{p}}   \qquad (|\x_0| \gg L),
\end{equation}
with an non-universal constant $C_{k,p} > 0$.  In particular, this
property ensures that the $L^2(\Omega)$-norm of $V_k^{(p)}$ is finite
for any $p>0$.

In turn, for $p = 0$, one gets
\begin{equation}  \label{eq:Jprime_p0}
\tilde{H}^\star_\infty(0|\x_0) = \begin{cases}  (|\x_0|/L)^{2-d}  \quad (d \geq 3), \cr 1  \hskip 19mm (d = 2). \end{cases}
\end{equation}
This is the hitting probability of the ball of radius $L$.  Due to the
recurrent nature of Brownian motion in two dimensions, this
probability is equal to $1$; in turn, it is smaller than $1$ for
Brownian motion in higher dimensions due to the possibility of escape
to infinity (transient motion).  As a consequence, each Steklov
eigenfunction $V_k^{(0)}(\x_0)$ is bounded by an explicit function
that decays via a power law as $|\x_0|\to \infty$ in dimensions $d
\geq 3$.  In turn, the above upper bound is useless for $d = 2$.

\section{Small-$p$ asymptotic behavior for bounded domains}
\label{sec:small-p}

For a bounded domain, the small-$p$ asymptotic behavior of the
eigenvalues were discussed earlier, see, e.g.,
\cite{Levitin08,Grebenkov19c}.  In \cite{Chaigneau24}, we used 
probabilistic arguments to get the asymptotic relation
(\ref{eq:muk_small_p_bounded}).  In this conventional setting, the
squared $L^2(\Omega)$-norm of $V_k^{(0)}$, given by $b_k$, is always
finite, ensuring the asymptotic behavior
(\ref{eq:muk_small_p_bounded}).  In this Appendix, we briefly discuss
a direct derivation of this result for a more general mixed Steklov
problem for bounded domains.

We consider a bounded domain $\Omega \subset \R^d$ whose boundary
$\pa$ is split into two parts, $\pa_S \cup \pa_D$, with Steklov and
Dirichlet boundary conditions on $\pa_S$ and $\pa_D$, respectively.
As in Sec. \ref{sec:identities}, we multiply $(p-\Delta) V_k^{(p)} =
0$ by $[V_k^{(p')}]^*$, multiply $(p'-\Delta) [V_k^{(p')}]^* = 0$ by
$V_k^{(p)}$, subtract these equations and integrate over $\Omega$ to
get
\begin{align} \nonumber
0 & = (p - p') \int\limits_{\Omega} V_k^{(p)} [V_k^{(p')}]^* 
- \int\limits_{\Omega} \biggl( [V_k^{(p')}]^* \Delta V_k^{(p)} - V_k^{(p)} \Delta [V_k^{(p')}]^*\biggr) \\
\label{eq:auxil8}
& = (p - p') \int\limits_{\Omega} V_k^{(p)} [V_k^{(p')}]^* 
- \int\limits_{\pa_S} \biggl([V_k^{(p')}]^* \partial_n V_k^{(p)} - V_k^{(p)} \partial_n [V_k^{(p')}]^*\biggr)
- \int\limits_{\pa_D} \biggl([V_k^{(p')}]^* \partial_n V_k^{(p)} - V_k^{(p)} \partial_n [V_k^{(p')}]^*\biggr) .
\end{align}
The last term vanishes due to the Dirichlet boundary condition on
$\pa_D$, yielding
\begin{equation}
(p' - p) \int\limits_{\Omega} V_k^{(p)} [V_k^{(p')}]^* = [\mu_k^{(p')}- \mu_k^{(p)}]
\int\limits_{\pa_S} v_k^{(p)} [v_k^{(p')}]^* . 
\end{equation}
From a general theory of elliptic equations in bounded domains
\cite{Kato}, the eigenvalues exhibit analytical dependence on $p$ (for
any $p > 0$), up to their suitable re-ordering to stay on the chosen
analytic branch (here we exclude the case of degenerate eigenvalues
that requires more subtle analysis, see also
\cite{Behrndt15,Bucur17}).
Setting $p' = p + \epsilon$ and taking the limit $\epsilon\to 0$, one
gets
\begin{equation}
\int\limits_\Omega |V_k^{(p)}|^2 = \partial_p \mu_k^{(p)}  \qquad (p \geq 0).
\end{equation}
As the $L^2(\Omega)$-norm of $V_k^{(0)}$ is finite for a bounded
domain, one retrieves Eq. (\ref{eq:muk_small_p_bounded}), with the
coefficients $b_k$ given by Eq. (\ref{eq:bk}).  Note that this
derivation remains valid for mixed Steklov-Neumann, Steklov-Robin and
Steklov-Robin-Dirichlet problems when the Dirichlet boundary condition
on $\pa_D$ is replaced by Neumann or Robin boundary condition.
Indeed, for all these boundary conditions, the last term in
Eq. (\ref{eq:auxil8}) still vanishes.

\section{Explicit computation for the exterior of a ball}
\label{sec:integral_ball}

In this Appendix, we focus on the exterior of a ball of radius $L$,
$E_L = \R^d \backslash \overline{B_L}$, and evaluate the integral
\begin{equation}
Q_n^{(p,L)} = \int\limits_{E_L} |V_{n,m}^{(p,L)}|^2 = \int\limits_{E_L} |\psi_{n,m}|^2 \, [g_n^{(p,L)}(r)]^2 
= \frac{1}{L^{d-1}} \int\limits_L^\infty dr \, r^{d-1} \, [g_n^{(p,L)}(r)]^2,
\end{equation}
where we used the factorization (\ref{eq:mun_ball_Rd}) of a Steklov
eigenfunction $V_{n,m}^{(p,L)}$ as the product of a spherical harmonic
$\psi_{n,m}$ and a radial function $g_n^{(p,L)}$.  Note that the
integral over angular coordinates yielded the factor $1/L^{d-1}$ due
to the normalization of spherical harmonics: $\int\nolimits_{\partial
B_L} |\psi_{n,m}|^2 = 1$.  Using the explicit form
(\ref{eq:gn_radial}) of the radial function, one has
\begin{equation}
Q_n^{(p,L)} = \frac{ \II_n(\alpha L)}{L^{d-1} [k_{n,d}(\alpha L)]^2 \alpha^d} \,,
\end{equation}
where $\alpha = \sqrt{p}$, and
\begin{equation}
\II_n(z) = \int\limits_z^\infty dx \, x^{d-1} \, [k_{n,d}(x)]^2 .
\end{equation}
The last integral can be found in a standard way by using the
differential equation (\ref{eq:knd_Bessel}) for $y(x) = k_{n,d}(x)$.
Multiplying this equation by $y(x)$ and integrating from $z$ to
$\infty$, one gets
\begin{equation} \label{eq:auxil5}
\int\limits_z^\infty dx\, x^{d-1} \biggl([y'(x)]^2 + [y(x)]^2 \bigl(1 + n(n+d-2)/x^2\bigr)\biggr) = - z^{d-1} y(z) y'(z).
\end{equation}  
In turn, multiplying Eq. (\ref{eq:knd_Bessel}) by $xy'(x)$ and
integrating from $z$ to $\infty$, one has
\begin{align} \nonumber
& \int\limits_z^\infty dx\, x^{d-1} \biggl((d-2)[y'(x)]^2 + [y(x)]^2 \bigl(d + (d-2) n(n+d-2)/x^2\bigr)\biggr) \\  \label{eq:auxil6}
& = z^d [y'(z)]^2 - z^d [y(z)]^2 (1 + n(n+d-2)/z^2).
\end{align}  
Multiplying Eq. (\ref{eq:auxil5}) by $d-2$ and subtracting from
Eq. (\ref{eq:auxil6}), one finds
\begin{equation}
\II_n(z) = \frac{z^{d-2}}{2} \biggl((d-2) z y(z) y'(z) + z^2 [y'(z)]^2 - y^2(z) [z^2 + n(n+d-2)] \biggr).
\end{equation}
Using the recurrence relation (\ref{eq:knd_rec}) we further simplify
this expression as
\begin{equation}  \label{eq:IIn}
\II_n(z) = \frac{z^{d-1} k_{n,d}^2(z)}{2} \biggl((2n+d-2)\frac{k_{n-1,d}(z)}{k_{n,d}(z)} 
+ z \biggl[\frac{k_{n-1,d}^2(z)}{k_{n,d}^2(z)} - 1\biggr]\biggr),
\end{equation}
and thus
\begin{equation}
Q_n^{(p,L)} = \frac{1}{2 \alpha} \biggl((2n+d-2)\frac{k_{n-1,d}(\alpha L)}{k_{n,d}(\alpha L)} 
+ \alpha L \biggl[\frac{k_{n-1,d}^2(\alpha L)}{k_{n,d}^2(\alpha L)} - 1\biggr]\biggr) \,.
\end{equation}

At the same time, one can easily check that the derivative of
$\mu_n^{(p,L)}$ given by Eq. (\ref{eq:mun_ball_Rd}) with respect to
$p$ yields the same expression.  We conclude that
\begin{equation}
Q_n^{(p,L)} = \partial_p \mu_n^{(p,L)} ,
\end{equation}
i.e., the squared $L^2(E_L)$-norm of $V_{n,m}^{(p,L)}$ is determined
by the derivative of the associated eigenvalue.  We therefore
retrieved the relation (\ref{eq:bk}) derived earlier for bounded
domains.

In the limit $p\to 0$, one uses the asymptotic behavior of the
modified Bessel function to get 
\begin{equation*}
k_{n,d}(z) \approx \Gamma(n-1+d/2) 2^{n-2+d/2} z^{-(n+d-2)} \qquad (z\to 0) 
\end{equation*}
so that $k_{n-1,d}(z)/k_{n,d}(z) \approx z/(2n+d-4)$ for $2n+d > 4$
and thus
\begin{equation}  \label{eq:Qn0}
Q_n^{(0,L)} = \begin{cases} L/(2n+d-4) \quad (n > 2-d/2), \cr +\infty \hskip 19.5mm  (n \leq 2-d/2). \end{cases}
\end{equation}
One can easily check that the divergence for $2n+d \leq 4$ occurs at
$n = 0$ for dimensions $d = 2,3,4$, and at $n = 1$ for dimension $d =
2$.

We use this result for evaluating the integral of a harmonic function
$u$ in the exterior of the ball.  Writing
\begin{equation}
u = \sum\limits_{n,m} V_{n,m}^{(0,L)} \bigl(u|_{\partial B_L}, \psi_{n,m}\bigr)_{L^2(\partial B_L)} ,
\end{equation}
one has
\begin{align*}
\int\limits_{E_L} |u|^2 &= \sum\limits_{n_1,m_1,n_2,m_2}  \bigl(u|_{\partial B_L}, \psi_{n_1,m_1}\bigr)^*_{L^2(\partial B_L)} 
 \bigl(u|_{\partial B_L}, \psi_{n_2,m_2}\bigr)_{L^2(\partial B_L)} 
\int\limits_{E_L} [V_{n_1,m_1}^{(0,L)}]^* V_{n_2,m_2}^{(0,L)} \\
& = \sum\limits_{n,m}  \bigl|\bigl(u|_{\partial B_L}, \psi_{n,m}\bigr)_{L^2(\partial B_L)}\bigr|^2 \int\limits_{E_L} |V_{n,m}^{(0,L)}|^2,
\end{align*}
where we used the orthogonality of spherical harmonics
$\{\psi_{n,m}\}$.  As a consequence, we obtain
\begin{equation}  \label{eq:u2_int}
\int\limits_{E_L} |u|^2 = \sum\limits_{n,m} Q_n^{(0,L)} 
\bigl|\bigl(u|_{\partial B_L}, \psi_{n,m}\bigr)_{L^2(\partial B_L)}\bigr|^2   \,.
\end{equation}
For instance, in three dimensions, if the integral of $u|_{\partial
B_L}$ over the sphere $\partial B_L$ is zero, the divergent term
$Q_0^{(0,L)}$ is excluded, and the integral (\ref{eq:u2_int}) is
finite.

\subsection{Numerical evaluation of the coefficient $b_k$}
\label{sec:bk}

To compute the coefficients $b_k$ from Eq. (\ref{eq:bk}), one can
split the unbounded domain $\Omega$ into $\Omega_L = \Omega \cap B_L$
and the exterior of the ball $B_L$, $E_L = \R^d \backslash
\overline{B_L}$ so that
\begin{equation}
b_k = \int\limits_{\Omega_L} |V_k^{(0)}|^2 + \int\limits_{E_L} |V_k^{(0)}|^2 .
\end{equation}
The first term can be calculated directly by using the FEM solution
inside the computational domain $\Omega_L$.  As discussed earlier, if
the integral of $v_k^{(0)}$ over $\pa$ is not zero, the coefficient
$b_k$ is infinite.  We therefore focus on the case when this integral
is zero and restrict the following discussion to three dimensions.
According to Eq. (\ref{eq:identity1_p0}), this implies that
$(V_k^{(0)}, 1)_{L^2(\partial B_L)} = 0$.  As $\psi_{0,0} =
1/\sqrt{4\pi L^2}$ is constant, one can exclude the term with $n = 0$
from the sum in Eq. (\ref{eq:u2_int}), yielding
\begin{equation}  \label{eq:V2_int_Linf}
\int\limits_{E_L} |V_k^{(0)}|^2 = \sum\limits_{n > 0} \sum\limits_{m=-n}^n \left|\bigl(\psi_{n,m}, V_k^{(0)}\bigr)_{L^2(\partial B_L)}\right|^2 
\frac{L}{2n-1} 
\end{equation}
in three dimensions.  

If we focus on the axisymmetric Steklov eigenfunctions, which do not
depend on the angle $\phi$, one can employ axisymmetric spherical
harmonics given by (normalized) Legendre polynomials:
\begin{equation}
\psi_{n,0}(\theta) = \sqrt{\frac{2n+1}{4\pi L^2}} \, P_n(\cos\theta).
\end{equation}
As a consequence, we have
\begin{equation}
\bigl(\psi_{n,0}, V_k^{(0)}\bigr)_{L^2(\partial B_L)} = L^2 \int\limits_0^{2\pi} d\phi \int\limits_0^\pi d\theta \, \sin\theta \,
\psi_{n,0}(\theta)\, V_k^{(0)}(L,\theta),
\end{equation} 
so that
\begin{equation}  \label{eq:V2_int_Linf2}
\int\limits_{E_L} |V_k^{(0)}|^2 = 2\pi L^3 \sum\limits_{n = 1}^\infty  
\frac{2n+1}{2n-1}  \left|\int\limits_0^\pi d\theta \, \sin\theta \,
P_n(\cos \theta)\, V_k^{(0)}(L,\theta)\right|^2 .
\end{equation}
In practice, this integral can be approximated by a Riemann sum which
is then evaluated over the nodes of the sphere $\partial B_L$, see
Appendix \ref{sec:numerics}.

In two dimensions, the $L^2(\Omega)$-norm of $V_k^{(0)}$ is finite
only if the first two terms with $n = 0$ and $n = 1$ are excluded
(i.e., their coefficients are strictly zero).  In this case, we get
\begin{align} \nonumber
\int\limits_{E_L} |V_k^{(0)}|^2 & = \sum\limits_{n \geq 2} \frac{L}{2n-2} \sum\limits_{m = \pm 1} 
\left|\bigl(\psi_{n,m}, V_k^{(0)}\bigr)_{L^2(\partial B_L)}\right|^2  \\   \label{eq:V2_int_Linf_2d}
& = \frac{L^2}{4\pi} \sum\limits_{n \geq 2} \frac{1}{n-1} \left\{
\left|\int\limits_0^{2\pi} d\theta \, e^{in\theta} \, V_k^{(0)}(L,\theta)\right|^2
+ \left|\int\limits_0^{2\pi} d\theta \, e^{-in\theta} \, V_k^{(0)}(L,\theta)\right|^2 \right\},
\end{align}
where we used $\psi_{n,m}(\theta) = e^{imn\theta}/\sqrt{2\pi L}$
(alternatively, one can use $\cos(n\theta)$ and $\sin(n\theta)$).

\section{Fourth identity}
\label{sec:proof}

Here we derive another identity, which can be used to evaluate and
analyze the terms in Eq. (\ref{eq:Wk}).  For this purpose, we consider
the Steklov eigenvalues $\hat{\mu}_{n,m}^{(p,L)}$ and eigenfunctions
$\hat{V}_{n,m}^{(p,L)}$ for the interior of a ball of radius $L$:
\begin{equation}
\hat{\mu}_n^{(p,L)} = \alpha \frac{I_{n-2+d/2}(\alpha L)}{I_{n-1+d/2}(\alpha L)} - \frac{n+d-2}{L} \,,
\qquad  \hat{V}_{n,m}^{(p,L)}(\x) = \psi_{n,m}(\theta,\cdots) \, \frac{i_{n,d}(\alpha r)}{i_{n,d}(\alpha L)}  \qquad (n=0,1,2,\ldots),
\end{equation}
where $\alpha = \sqrt{p}$, $i_{n,d}(z) = z^{1-d/2} I_{n-1+d/2}(z)$ is
an extension of the modified Bessel function $I_\nu(z)$ of the first
kind, and we used the recurrence relation 
\begin{equation}
i'_{n,d}(z) = i_{n-1,d}(z) - \frac{n+d-2}{z} i_{n,d}(z).
\end{equation}
In the limit $p\to 0$, one gets $\hat{\mu}_n^{(0,L)} = n/L$ and
$\hat{V}_{n,m}^{(0,L)}(\x) = \psi_{n,m}(\theta,\cdots) (r/L)^n$; in
particular, the principal eigenvalue is zero, as for any interior
problem.

Using the eigenfunctions $\hat{V}_{n,m}^{(p,L)}$, we can express the
integrals of $V_k^{(p)}$ multiplied by $\psi_{n,m}$ on the sphere
$\partial B_L$.  Indeed, multiplying $(p-\Delta)V_k^{(p)} = 0$ by
$\hat{V}_{n,m}^{(p,L)}$, multiplying $(p-\Delta)\hat{V}_{n,m}^{(p,L)}
= 0$ by $V_k^{(p)}$, subtracting and integrating over $\Omega_L$, one
gets
\begin{equation*}
0 = \int\limits_{\partial B_L} \biggl(\hat{V}_{n,m}^{(p,L)} \underbrace{\partial_n V_k^{(p)}}_{-\M_p^L V_k^{(p)}} 
- V_k^{(p)} \underbrace{\partial_n \hat{V}_{n,m}^{(p,L)}}_{=\hat{\mu}_n^{(p,L)} \hat{V}_{n,m}^{(p,L)}} \biggr)
+ \int\limits_{\pa} \biggl(\hat{V}_{n,m}^{(p,L)} \underbrace{\partial_n V_k^{(p)}}_{=\mu_k^{(p)} V_k^{(p)}} 
- V_k^{(p)} \partial_n \hat{V}_{n,m}^{(p,L)}\biggr)
\end{equation*}
(note that the integer index $n$ should not be confusing with the
normal derivative $\partial_n$).  Using
$\hat{V}_{n,m}^{(p,L)}|_{\partial B_L} = \psi_{n,m}$ and the
self-adjointness of $\M_p^L$, we find the fourth identity
\begin{equation}  \label{eq:fourth_identity}
\bigl(\mu_n^{(p,L)} + \hat{\mu}_n^{(p,L)}\bigr) \int\limits_{\partial B_L} \psi_{n,m} V_k^{(p)} 
= \int\limits_{\pa} v_k^{(p)} \biggl(\mu_k^{(p)} \hat{V}_{n,m}^{(p,L)} - \partial_n \hat{V}_{n,m}^{(p,L)}\biggr).
\end{equation}

Note also that using the inequality \cite{Ifantis90} 
\begin{equation}
\frac{I_\nu(z)}{I_{\nu+1}(z)} < 1 + \frac{2(\nu+1)}{z}   \qquad (z > 0, ~ \nu > -1)
\end{equation}
and setting $\nu = n-2+d/2$, we conclude that
\begin{equation}
\hat{\mu}_n^{(p,L)} - \hat{\mu}_n^{(0,L)} = \sqrt{p} \frac{I_{n-2+d/2}(\sqrt{p} L)}{I_{n-1+d/2}(\sqrt{p} L)} - \frac{2n+d-2}{L} \leq \sqrt{p}
\qquad (p \geq 0)
\end{equation}
for any $n = 0,1,2,\ldots$ and $d \geq 2$.

\section{Numerical method for solving the exterior Steklov problem}
\label{sec:numerics}

Various numerical methods have been developed for solving the Steklov
problem in bounded domains
\cite{Andreev04,Yang09,Bi11,Li11,Li13,Xie13,Bi16,Akhmetgaliyev17,Alhejaili19,Bruno20,Chen20}.
For instance, one can use a weak formulation of the boundary value
problem (\ref{eq:Steklov}) and basis functions on a mesh of a given
bounded domain to get a matrix representation of the
Dirichlet-to-Neumann operator.  A numerical diagonalization of such a
truncated matrix allows one to approximate a number of eigenvalues
$\mu_k^{(p)}$ and to construct the associated eigenfunctions
$v_k^{(p)}$ on $\pa$ and the Steklov eigenfunctions $V_k^{(p)}$ on
$\Omega$.  In \cite{Chaigneau24}, this finite-element method (FEM) was
also adapted for solving mixed Steklov problems in bounded domains
when the Steklov boundary condition $\partial_n V_k^{(p)} =
\mu_k^{(p)} V_k^{(p)}$ on a subset $\Gamma$ of the boundary $\pa$ is
completed by Dirichlet, Neumann or Robin boundary condition on the
remaining part $\pa\backslash \Gamma$.  In this Appendix, we discuss
its extension to exterior domains.

\subsection{Transparent boundary condition}

We look for the solution of the exterior boundary value problem:
\begin{equation}
(p - \Delta) u = 0 \quad \textrm{in}~ \Omega, \qquad u|_{\pa} = f, \qquad u|_{|\x|\to \infty} =0,
\end{equation}
for a given function $f$ on $\pa$ and $p \geq 0$.  The main difficulty
for implementing a FEM is that the domain $\Omega$ is unbounded.

Let $B_L$ be the ball of radius $L$, centered at the origin, that
includes $\pa$.  We introduce the bounded domain $\Omega_L = \Omega
\cap B_L$, which lies between the inner boundary $\pa$ (with Steklov
condition) and the outer boundary $\partial B_L$.  As $u$ satisfies
the modified Helmholtz equation in $\Omega$, it is an analytic
function away from the boundary $\pa$; in particular, $u$ and its
derivative are continuous on $\partial B_L$:
\begin{equation}
u|_{\partial B_L^-} = u|_{\partial B_L^+},  \qquad
(\partial_n u)|_{\partial B_L^-} = - (\partial_n u)|_{\partial B_L^+}, 
\end{equation}
where $\partial B_L^{-}$ and $\partial B_L^{+}$ denote the approach to
the sphere $\partial B_L$ from the inner and the outer sides of
$\partial B_L$, respectively.  By introducing an auxiliary
Dirichlet-to-Neumann operator $\M_p^L$ in the exterior of the ball
$B_L$, one gets
\begin{equation}
(\partial_n u)|_{\partial B_L^+} = \M_p^L u|_{\partial B_L^+} \quad \Rightarrow \quad 
(\partial_n u)|_{\partial B_L^-} = - \M_p^L u|_{\partial B_L^-}.
\end{equation}
In other words, one can focus on the solution of the following
boundary value problem in the bounded domain $\Omega_L$:
\begin{equation}   \label{prob:helm}
\left\{ \begin{array}{r l l}
(p - \Delta) u(\x)& = 0 &\quad (\x \in \Omega_L), \\
u(\x)& = f(\x)& \quad (\x \in \pa),\\  
\partial_n u(\x) &= - \M_p^L u(\x)& \quad (\x \in \partial B_L) ,
\end{array}  \right.
\end{equation}
where we wrote $\partial B_L$ instead of $\partial B_L^{-}$ for
brevity.  In this way, the original problem in the unbounded domain
$\Omega$ is reduced to an equivalent problem in the bounded domain
$\Omega_L$ with a Robin-like boundary condition on $\partial B_L$,
known as transparent boundary condition (TBC) (see
\cite{Keller89,Givoli91,Hagstrom99,Givoli04,Achdou07,Antoine08} and
references therein).  In particular, one can now apply the FEM to
compute the eigenvalues and eigenfunctions of the Dirichlet-to-Neumann
operator $\M_p$ in $\Omega_L$, which are identical to those in the
original domain $\Omega$.  A similar approach to electromagnetic and
acoustic waves employed so-called perfectly matched layer condition
\cite{Berenger94,Collino98,Becache04,Becache06,Kalvin12} (see
also \cite{Levitin08b}).

The choice of a ball $B_L$ as an enclosing domain has the advantage
that the spectral properties of the auxiliary Dirichlet-to-Neumann
operator $\M_p^L$ are known explicitly (see Sec. \ref{sec:ball}).  In
fact, the action of $\M_p^L$ onto a function $f$ on $\partial B_L$ can
be written as
\begin{equation}  \label{eq:MpL_f}
\M_p^L f = \sum\limits_{n,m} \mu_n^{(p,L)} \psi_{n,m} \int\limits_{\partial B_L}  \psi_{n,m}^*\, f ,
\end{equation}
where $\mu_n^{(p,L)}$ and $\psi_{n,m}$ are the eigenvalues and
eigenfunctions of $\M_p^L$.

A practical implementation of the transparent boundary condition on
$\partial B_L$ is straightforward.  The infinite sum in
Eq. (\ref{eq:MpL_f}) has to be truncated by considering only the
eigenfunctions $\psi_{n,m}$ whose eigenvalues $\mu_n^{(p,L)}$ do not
exceed a chosen threshold $\mu$.  We enumerate a finite number of such
eigenfunctions by a single index $j$ ranging from $1$ to $J$.  
For instance, in two dimensions, one can fix the truncation order
$\nmax$ and then take the first $J = 2\nmax + 1$ eigenmodes, given
that $\psi_{n,m} = e^{inm\theta}/\sqrt{2\pi L}$, with $m = \pm 1$ and
$n = 0,1,\ldots, \nmax$.  In three dimensions, one would in general
have $J = (\nmax+1)^2$ spherical harmonics $\psi_{n,m}$ with $n \leq
\nmax$.  However, it is possible to make $J$ proportional to $\nmax$
for axisymmetric domains, as described in the next subsection.
For the set $\{\x_1,\ldots,\x_N\}$ of mesh points on $\partial B_L$,
we introduce the matrix $\mathbf{V}$ of size $N \times J$ with the
elements $\mathbf{V}_{i,j} = \psi_{n_j,m_j}(\x_i)$, and the diagonal
matrix $\mathbf{m}$ of size $J \times J$ with the elements
$\mathbf{m}_{i,j} = \delta_{i,j} \mu_{n_j}^{(p,L)}$.  If $\mathbf{f}_i
= f(\x_i)$ are the elements of a vector $\mathbf{f}$ representing the
values of a given function $f$ at points $\{\x_i\}$, one gets
\begin{equation}    \label{eq:PMC_1}
(\partial_n u)(\x_i) = - (\M_p^L f)(\x_i) \approx - \sum\limits_{j=1}^J  \mu_{n_j}^{(p,L)} 
\mathbf{V}_{i,j} \sum\limits_{k=1}^N ds_k \mathbf{V}_{k,j}^* \mathbf{f}_k  
= - [\mathbf{V} \mathbf{m} \mathbf{V}^\dagger \mathbf{S} \mathbf{f}]_i ,
\end{equation}
where $[\mathbf{V}^\dagger]_{i,j} = \mathbf{V}^*_{j,i}$, and
$\mathbf{S}_{j,j'} = \delta_{j,j'} ds_j$ are the elements of the
diagonal matrix $\mathbf{S}$ of size $N\times N$ formed by the surface
elements $ds_j$ (e.g., in two dimensions, $ds_j$ is the distance
between two neighboring points $\x_j$ and $\x_{j+1}$ on $\partial
B_L$).  In other words, the transparent boundary condition turns out
to be equivalent to a generalized Robin boundary condition when the
normal derivative of $u$ is proportional to an integral kernel acting
on $u$ on $\partial B_L$.  Such a boundary condition was implemented
in \cite{Chaigneau24} through a matrix $\mathbf{Q}$.  Setting
$\mathbf{Q}=- \mathbf{V} \mathbf{m} \mathbf{V}^\dagger \mathbf{S}$, we
can thus incorporate the transparent boundary condition on $\partial
B_L$.

A potential limitation of this implementation is that the eigenvalues
$\mu_n^{(p,L)}$ grow with $n$ so that the approximation
(\ref{eq:PMC_1}) might be sensitive to truncation.  To resolve this
issue, one can implement an equivalent condition:
\begin{equation}  \label{eq:PML_bis}
- (\M_p^L)^{-1} (\partial_n u)|_{\partial B_L} = u|_{\partial B_L} ,
\end{equation}
that is applicable when $\mu_0^{(p,L)} > 0$ and admits a similar
matrix representation.  However, for all examples considered in the
paper, the representation (\ref{eq:PMC_1}) did not manifest any
numerical instability so that we did not need to resort to the
alternative form (\ref{eq:PML_bis}).

Once the Steklov eigenfunctions $V_k^{(p)}$ are constructed inside the
computational domain $\Omega_L$ via the above FEM, their extension
outside $\Omega_L$ can be easily performed.  In fact, the restriction
of $V_k^{(p)}$ onto $\partial B_L$ uniquely determines its extension
to $E_L = \R^d \backslash \overline{B_L}$ as
\begin{equation}
V_k^{(p)}(\y) = \sum\limits_{n,m} V_{n,m}^{(p,L)}(\y) \int\limits_{\partial B_L} \psi_{n,m}^* V_k^{(p)}  \qquad (|\y| \geq L).
\end{equation}
As previously, the infinite sum has to be truncated to a finite number
of terms, whereas the integral over the sphere $\partial B_L$ is
approximated by a Riemann sum on mesh points lying on $\partial B_L$.
As a consequence, the evaluation of $V_k^{(p)}$ on a set of points
$\y_1,\ldots,\y_M$ in $E_L$ can be expressed in a matrix form,
allowing for a rapid computation.

\subsection{FEM for three-dimensional axisymmetric domains}

The FEM that was originally developed in \cite{Chaigneau24} for planar
domains, is directly applicable in higher dimensions.  A standard
practical limitation of the FEM is that a representation of the
modified Helmholtz equation on a higher-dimensional mesh would involve
much larger matrices and thus require substantially higher
computational resources.  For this reason, we focus here on
three-dimensional axisymmetric domains, for which the computation can
be reduced to planar domains and is thus much simpler and faster.

\begin{figure}
\includegraphics[width=\linewidth]{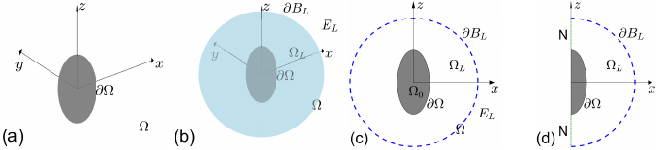}
\caption{
Axisymmetric domains in three dimensions.  (a) the exterior $\Omega =
\R^3\backslash \Omega_0$ of a prolate spheroid $\Omega_0$ (in gray);
(b) the exterior domain $\Omega$ is split into a bounded computational
domain $\Omega_L = \Omega \cap B_L$ (in light blue) and the exterior
$E_L$ of a ball $B_L$; the Steklov condition is imposed on the
boundary $\pa$, while the transparent boundary condition is imposed on
the sphere $\partial B_L$; (c) projection of the domain $\Omega$ onto
the $xz$ plane; (d) half of this projection that corresponds to $\phi
= 0$ in cylindrical coordinates $(r,z,\phi)$; one can thus associate
$r = x$, employ Neumann boundary condition (N) on two vertical
segments, and solve the mixed Steklov problem in the bounded
computational domain $\Omega_L$.  }
\label{fig:scheme}
\end{figure}

In practice, we assume the rotational symmetry around the $z$-axis
(see Fig. \ref{fig:scheme}) and employ the cylindrical coordinates
$(r,z,\phi)$, in which the Laplace operator reads
\begin{equation}\label{eq:cylindlap}
\Delta = \frac{1}{r}\partial_r r \partial_r + \partial_z^2 + \frac{1}{r^2} \partial_{\phi}^2,
\end{equation}
where $r=\sqrt{x^2+y^2}$.  If a given function $f$ on the boundary
$\pa$ of an axisymmetric domain does not depend on $\phi$, the last
term in Eq. (\ref{eq:cylindlap}) can be dropped, and the modified
Helmholtz equation can be written in a matrix form as
\begin{equation}  \label{eq:helm_2d}
-\nabla \cdot (c\nabla u) + pru= 0 ,
\end{equation}
where $c$ is a $2\times2$ diagonal matrix whose elements are given by
$r$.  As a consequence, the original three-dimensional problem in
$\Omega$ is reduced to solving Eq. (\ref{eq:helm_2d}) in a planar
domain that is the projection of $\Omega$ onto the $xz$ plane (i.e.,
by setting $\phi = 0$ and $\phi = \pi$).  Moreover, we further reduce
this planar computational domain by considering only its half that
corresponds to $\phi = 0$.  To ensure the axial symmetry, one imposes
an additional Neumann boundary condition at the vertical segment at $r
= x = 0$.  In this way, one can directly apply the FEM, as described
in \cite{Chaigneau24}.  This computation can be easily extended to
axisymmetric domains in higher dimensions.

\subsection{Computation of non-axisymmetric eigenfunctions}

For three-dimensional domains, we only presented axisymmetric Steklov
eigenfunctions which do not depend on the angular coordinate $\phi$.
To get a numerical solution in the general case, one needs to solve
the modified Helmholtz equation in cylindrical coordinates
$(r,z,\phi)$:
\begin{equation}
(\Delta - p) u = \biggl(\frac{1}{r} \partial_r r \partial_r + \partial_z^2 + \frac{1}{r^2} \partial_\phi^2 - p\biggr) u = 0.
\end{equation}
The axial symmetry of the domain $\Omega$ (which is still
axisymmetric) allows one to search a solution in the form $u(r,z,\phi)
= e^{im\phi} U_m(r,z)$, with an integer $m$, where $U_m(r,z)$
satisfies
\begin{equation}
\biggl(\partial_r r \partial_r + r\partial_z^2 - \frac{m^2}{r} - pr \biggr) U_m = 0.
\end{equation}
Following the standard FEM procedure of representing the solution on
the basis of functions $\varphi_j$ (see \cite{Chaigneau24} for
details), one gets in a matrix form
\begin{equation}
\mathbf{U} \bigl[\mathbf{K} + p \mathbf{M} + m^2 \hat{\mathbf{M}}\bigr] = \mathbf{F},
\end{equation}
where the vectors $\mathbf{U}$ and $\mathbf{F}$ represent the solution
$U_m$ and the boundary condition on basis functions, $\mathbf{K}$ and
$\mathbf{M}$ are the usual stiff and mass matrices, whereas the matrix
$\hat{\mathbf{M}}$ has the matrix elements
\begin{equation}
\hat{\mathbf{M}}_{j,k} = \int\limits_{\Omega} \frac{1}{r} \varphi_j \varphi_k .
\end{equation}
In other words, one needs to replace the matrix $p\mathbf{M}$ by the
matrix $p\mathbf{M} + m^2 \hat{\mathbf{M}}$ in the original
construction of the matrix representation of the Dirichlet-to-Neumann
operator $\M_p$ \cite{Chaigneau24}.  In the special case $m = 0$, the
diagonalization of the matrix representing $\M_p$ yields the
axisymmetric eigenfunctions and the associated eigenvalues of $\M_p$,
as considered in the main text.  In the same vein, setting $m =
1,2,\ldots$ allows one to construct other matrix representations,
whereas their diagonalization will result in non-axisymmetric
eigenfunctions and associated eigenvalues of $\M_p$.

\subsection{Validation}

\begin{figure}
\centering
\includegraphics[width=0.35\linewidth]{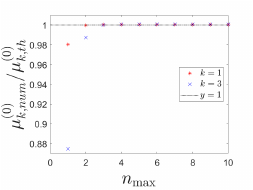}
\caption{
Ratio between numerical and analytical eigenvalues $\mu_1^{(0)}$ and
$\mu_3^{(0)}$ for the exterior of the unit disk as a function of
$\nmax$, with $\hmax = 0.025$. }
\label{fig:mu-vs-smallp0}
\end{figure}

We validate the proposed FEM with the transparent boundary condition
by solving the Steklov problem for the exterior of a disk, for which
the eigenvalues are known explicitly (see Sec. \ref{sec:ball}).  To
limit of the effect of the disk symmetry, the center of the disk is
shifted from the origin.  In practice, we consider a disk of radius
$R=1$ centered at $(0, 0.25)$ and surrounded by a circle $\partial
B_L$ of radius $L=2$ centered at the origin.  The exact eigenvalues
for this exterior problem are given by $\mu_n^{(p)} = -\sqrt{p}
\frac{K_n'(R \sqrt{p})}{K_n(R\sqrt{p})}$, which reduce to $\mu_n^{(0)}
= n/R$ at $p = 0$.  The rotational symmetry of the domain implies that
the eigenfunctions are the Fourier harmonics $e^{m
in\theta}/\sqrt{2\pi R}$ (with $m = \pm 1$), which do not depend on
$p$.  Table \ref{tab:2D_p0} compares the first 11 eigenvalues to those
obtained by the FEM on two meshes with different maximal mesh sizes
$\hmax$, for $p = 0$ and $p = 1$.  The root mean squared errors of the
associated eigenfunctions are also shown.  In turn,
Fig. \ref{fig:mu-vs-smallp0} illustrates how fast the numerical
eigenvalues converge to the theoretical value as the truncation order
$\nmax$ increases.

\begin{table}[t!]
\centering
\begin{tabular}{|c|c|c|c|c|c||c|c|c|c|c|} \hline 
& \multicolumn{5}{c||}{$p=0$} & \multicolumn{5}{c|}{$p=1$} \\ \hline
& \multicolumn{3}{c|}{Eigenvalues} & \multicolumn{2}{c||}{RMSE} 
& \multicolumn{3}{c|}{Eigenvalues} & \multicolumn{2}{c|}{RMSE} \\ \hline
Index $k$  & Exact & $\hmax = 0.025$  & $\hmax = 0.01$ & $\hmax = 0.025$  & $\hmax = 0.01$ 
& Exact & $\hmax = 0.025$ & $\hmax = 0.01$ & $\hmax = 0.025$ & $\hmax = 0.01$ \\  \hline 
0 & 0 & $1.25\times10^{-9}$ & $2.78\times10^{-10}$ & 0.0014 & 0.0001
  & 1.4296 & 1.4302 & 1.4293 & 0.007 & 0.0004 \\
1 & 1 & 1.0007 & 0.9996 & 0.0022 & 0.0013
  & 1.6995 & 1.7002 & 1.6992 & 0.0059 & 0.0032\\
2 & 1 & 1.0007 & 0.9996 & 0.0022 & 0.0013
  & 1.6995 & 1.7003 & 1.6993 & 0.0059 & 0.0032\\
3 & 2 & 2.0017 & 2.0000 & 0.0052 & 0.0031
  & 2.3704 & 2.3723 & 2.3705 & 0.0204 & 0.0032\\
4 & 2 & 2.0017 & 2.0000 & 0.0052 & 0.0031
  & 2.3704 & 2.3723 & 2.3705 & 0.0204 & 0.0031\\
5 & 3 & 3.0042 & 3.0005 & 0.0088 & 0.0060
  & 3.2288 & 3.2333 & 3.2294 & 0.0235 & 0.0081\\
6 & 3 & 3.0042 & 3.0005 & 0.0088 & 0.0060
  & 3.2288 & 3.2333 & 3.2294 & 0.0235 & 0.0081\\
7 & 4 & 4.0090 & 4.0013 & 0.0197 & 0.0079
  & 4.1605 & 4.1699 & 4.1619 & 0.0239 & 0.0050\\
8 & 4 & 4.0091& 4.0013 & 0.0197  & 0.0079
  & 4.1605 & 4.1700 & 4.1619 & 0.0219 & 0.0050\\
9 & 5 & 5.0167 & 5.0026 & 0.0197 & 0.0111
  & 5.1225 & 5.1397 & 5.1252 & 0.0248 & 0.0103\\
10 & 5 & 5.0167 & 5.0026 & 0.0284 & 0.0111
  & 5.1225 & 5.1397 & 5.1252 & 0.0248 & 0.0103\\ \hline
\end{tabular}
\caption{
List of the first 11 eigenvalues $\mu_k^{(0)}$ (columns 2-4) and root
mean squared errors (columns 5-6) for the associated eigenfunctions
$v_k^{(0)}$ for the exterior of the unit disk.  Similar information
for $p = 1$ is provided in columns 7-11.  We used $L = 2$ and $\nmax =
30$. }
\label{tab:2D_p0}
\end{table}

In three dimensions, we consider a sphere of radius $R=1$ centered at
$(0, 0, 0.25)$, surrounded by a sphere of radius $L=2$ centered at the
origin.  The comparison between exact and numerically obtained
eigenvalues and eigenfunctions was very similar and thus not shown.

\end{document}